\documentclass{aa}  

\usepackage{natbib}
\usepackage{graphicx}
\usepackage{txfonts}

\usepackage{xcolor}
\usepackage{tikz}
\usetikzlibrary{bayesnet}

\newcommand{\de}{\text{d}}
\newcommand{\Msun}{\text{M}_{\odot}}
\newcommand{\kmsec}{\text{km}\text{s}^{-1}}
\newcommand{\kmsecsq}{\text{km}^2\text{s}^{-2}}
\newcommand{\kpc}{\text{kpc}}
\newcommand{\pc}{\text{pc}} 
\newcommand{\yr}{\text{yr}}
\newcommand{\Msunppcc}{\Msun\pc^{-3}}
\newcommand{\Msunppcsquare}{\Msun\pc^{-2}}

\newcommand{\mas}{\text{mas}}

\newcommand{\masyr}{\text{mas}~\text{yr}^{-1}}

\newcommand{\popp}{\boldsymbol{\Psi}}
\newcommand{\objp}{\boldsymbol{\psi}}

\newcommand{\data}{\boldsymbol{\hat{d}}}

\newcommand{\zlim}{Z_{\text{lim.}}}

\usepackage{hyperref}
\begin{document}

   \title{Weighing the Galactic disk in sub-regions of the solar neighbourhood using \emph{Gaia} DR2}

   \author{A. Widmark
          \inst{1}
          \and
          P.F. de Salas
          \inst{2}
          \and
          G. Monari\inst{3}
          }

   \institute{Dark Cosmology Centre, Niels Bohr Institute, University of Copenhagen, Jagtvej 128, 2200 Copenhagen N, Denmark\\
   \email{axel.widmark@nbi.ku.dk}
   \and
   The Oskar Klein Centre for Cosmoparticle Physics, Department of Physics, Stockholm University, AlbaNova, 10691 Stockholm, Sweden
   \and
   Universit\'e de Strasbourg, CNRS UMR 7550, Observatoire astronomique de Strasbourg, 11 rue de l'Universit\'e, 67000 Strasbourg, France
    }

   \date{Received Month 11, 2020; accepted Month XX, XXXX}

 
  \abstract
   {}
   {We infer the gravitational potential of the Galactic disk by analysing the phase-space densities of 120 stellar samples in 40 spatially separate sub-regions of the solar neighbourhood, using \emph{Gaia}'s second data release (DR2), in order to quantify spatially dependent systematic effects that bias this type of measurement.}
   {The gravitational potential was inferred under the assumption of a steady state in the framework of a Bayesian hierarchical model. We performed a joint fit of our stellar tracers' three-dimensional velocity distribution, while fully accounting for the astrometric uncertainties of all stars as well as dust extinction, and we also masked angular areas of known open clusters. The inferred gravitational potential is compared, post-inference, to a model for the baryonic matter and halo dark matter components.}
   {We see an unexpected but clear trend for all 40 spatially separate sub-regions: Compared to the potential derived from the baryonic model, the inferred gravitational potential is significantly steeper close to the Galactic mid-plane ($\lesssim 60~\pc$), but flattens such that the two agree well at greater distances ($\sim 400~\pc$). The inferred potential implies a total matter density distribution that is highly concentrated to the Galactic mid-plane and decays quickly with height. We see a dependence on the Galactic radius that is consistent with a disk scale length of a few kiloparsecs. Apart from this, there are discrepancies between stellar samples, implying spatially dependent systematic effects which are, at least in part, explained by substructures in the phase-space distributions.}
   {In terms of the inferred matter density distribution, the very low matter density that is inferred at greater heights ($\gtrsim 300~\pc$) is inconsistent with the observed scale height and matter distribution of the stellar disk, which cannot be explained by a misunderstood density of cold gas or other hidden mass. Our interpretation is that these results must be biased by a time-varying phase-space structure, possibly a breathing mode, that is large enough to affect all stellar samples in the same manner.}

   \keywords{Galaxy: kinematics and dynamics -- Galaxy: disc -- solar neighborhood -- Astrometry}

   \maketitle
%

\section{Introduction}\label{sec:intro}
Dynamical mass measurements of the solar neighbourhood and Galactic disk are of great importance for constraining properties of the Milky Way \citep{1998MNRAS.294..429D, Klypin:2001xu, Widrow:2008yg,2010A&A...509A..25W,McMillan:2011wd,2014ApJ...794...59K,2017MNRAS.465..798C,2017MNRAS.465...76M,2020MNRAS.494.6001N,2020MNRAS.494.4291C,2020ApJ...894...10L}, and this has implications for the dark sector, for direct and indirect dark matter detection experiments \citep{Jungman:1995df,2015PrPNP..85....1K}, as well as for dark substructures \citep{10.1111/j.1365-2966.2008.13643.x,0004-637X-703-2-2275,Fan:2013tia,2014MNRAS.444..515R}. Dynamical mass measurements are often performed under the assumption of a steady state by isolating a stellar tracer population and fitting its phase-space distribution or through Jeans analysis \citep{Kapteyn1922,Oort1932,1984ApJ...287..926B,1984ApJ...276..169B,KuijkenGilmore1989b,KuijkenGilmore1989c,KuijkenGilmore1989a,KuijkenGilmore1991,Creze1998,HolmbergFlynn2000,Bienayme:2005py,doi:10.1111/j.1365-2966.2012.21608.x,0004-637X-756-1-89,Zhang_2013,2014A&A...571A..92B,WidmarkMonari,Schutz:2017tfp,Buch:2018qdr,Widmark2019,2020A&A...643A..75S,2020MNRAS.495.4828G}. With the depth and precision of the astrometric \emph{Gaia} mission \citep{2016A&A...595A...1G}, which had its second data release (DR2) in April 2018 \citep{2018A&A...616A...1G}, it is possible to fit more complicated dynamical models than ever before, calling for more sophisticated statistical methods and modelling techniques.

The solar neighbourhood and the Galactic mid-plane is believed to be dominated by baryons with roughly equal matter densities of stars and gas \citep{Flynn:2006tm,2015ApJ...814...13M,0004-637X-829-2-126}. Precise dynamical mass measurements of the Galactic disk can constrain the distribution of baryons, supplementing other observations, as well as the local density of dark matter. In the standard scenario, the cold dark matter halo would constitute about 10 \% of the mid-plane matter density \citep{2020A&A...643A..75S,2020MNRAS.495.4828G}. A more exotic possibility has been proposed by \cite{Fan:2013tia,Fan:2013yva}, whereby a thin dark disk, co-planar with and embedded in the stellar disk of the Milky Way, would form from a dark matter sub-component with strong dissipative self-interactions. The possibility of a thin dark disk was constrained by \cite{Kramer:2016dqu} using interstellar gas, by \cite{Caputo:2017zqh} using binary pulsars, and by \cite{Schutz:2017tfp} and \cite{Buch:2018qdr} using \emph{Gaia} observations. The studies based on \emph{Gaia} set the most stringent upper bounds to such a surface density surplus, of approximately $7~\Msunppcsquare$. However, a third \emph{Gaia} based study by \cite{Widmark2019} infers an over-density in the Galactic plane that is on par with these upper bounds.

This paper is largely a follow-up on the analysis made in \cite{Widmark2019}. Following the approach of that work, we formulated our statistical model in the framework of a Bayesian hierarchical model and utilised the six-dimensional phase-space information of all stars in our stellar samples, while fully accounting for the astrometric uncertainties of all stars. We performed a joint fit of the gravitational potential and the full three-dimensional velocity distribution (rather than just the vertical velocities, as in previous work). We also accounted for dust extinction and masked the angular areas where open clusters are known to be present. We divided the solar neighbourhood into 120 stellar samples, using 40 spatially separate area cells in the directions parallel to the Galactic plane, and three different cuts in the vertical direction (200, 300, and 400 pc above and below the Sun). The main reason for this division was to quantify possible systematic effects, which would reveal themselves if they affect the stellar samples differently, depending on their spatial position. Such effects could be associated with substructures and time-variations of the phase-space distribution, breaking the steady state assumption. Many such features have already been revealed by \emph{Gaia}, such as moving groups in the solar neighbourhood and phase-space spirals and ridges \citep{gaia_kinematics,wrinkles}. It is an open question to what extent such time-varying features can affect this type of local dynamical mass measurement. The main focus of this work is to quantify such biases and determine whether they have a spatial dependence.

This article is structured in the following way. In Sect.~\ref{sec:galacticmodel}, we present our model for the dynamics of the solar neighbourhood as well as the expected gravitational potential coming from baryonic observations. In Sect.~\ref{sec:data}, we discuss the data of \emph{Gaia} DR2 and explain how the stellar samples are constructed. The statistical model is present in Sect.~\ref{sec:statisticalmodel}, and the inferred results are found in in Sect.~\ref{sec:results}. In Sects.~\ref{sec:discussion} and \ref{sec:conclusion}, we discuss and conclude.

\section{Galactic model}\label{sec:galacticmodel}

\subsection{Coordinate system}\label{sec:coordinate_system}
In order to describe the phase-space density of stars in the local neighbourhood, we use the following system of coordinates. The spatial coordinates $\boldsymbol{X} \equiv \{X,Y,Z\}$ denote the position of a star as seen from the Sun, where positive $X$ corresponds to the direction of the Galactic centre, positive $Y$ corresponds to the direction of Galactic rotation, and positive $Z$ corresponds to the direction of Galactic north. Their respective time derivatives correspond to velocities $\boldsymbol{V} \equiv \{U,V,W\}$, whose origin is that of the solar rest frame.

Furthermore, the quantities $z$ and $w$ correspond to position and velocity as seen from the rest-frame of the Galactic mid-plane. They are equal to
\begin{equation}
\begin{split}
    z & = Z + Z_\odot, \\
    w & = W + W_\odot,
\end{split}
\end{equation}
where $Z_\odot$ is the vertical position of the Sun with respect to the Galactic mid-plane and $W_\odot$ is the vertical velocity of the Sun in the rest-frame of the Galactic disk.

\subsection{Gravitational potential}\label{sec:gravitational_potential}
In this work we considered stellar tracer populations in the solar neighbourhood, out to a distance of a few hundred parsecs from the Sun. We constructed a total of 120 stellar samples (see Sect.~\ref{sec:samplecuts} for a detailed description) whose spatial volumes are elongated in the direction of $Z$ (extending 400--800 pc), and comparatively narrow in $X$ and $Y$ (50--100 pc). In these column-like volumes, we made the assumption that the gravitational potential only depends on $Z$ and is invariant with respect to $X$ and $Y$.

The gravitational potential was modeled as
\begin{equation}\label{eq:gravitational_potential}
    \Phi(z) = 4 \pi G
    \sum_{h} \; \rho_h\, \text{log}\Bigg[ \text{cosh} \Bigg( \frac{z}{h~\pc} \Bigg) \Bigg] \, (h~\pc)^2,
\end{equation}
where $h$ iterates over the numbers $\{40,\, 80,\, 160,\, 320\}$.

Given the one-dimensional Poisson equation,
\begin{equation}\label{eq:Poisson}
	\frac{\partial^2\Phi(z)}{\partial z^2} = 4\pi G \rho(z),
\end{equation}
the four parameters $\rho_{\{40, 80, 160, 320\}}$ correspond to a sum of matter density components with scale heights of 40, 80, 160, and 320 pc respectively, according to
\begin{equation}\label{eq:density_distribution}
\rho(z) = \sum_{h} \; \rho_h \, \text{sech}^2 \Bigg( \frac{z}{h~\pc} \Bigg).
\end{equation}
This model form assumes mirror symmetry of the Galactic plane and that the matter density decreases monotonically with distance from the Galactic mid-plane, but is otherwise quite free to vary in shape.

The one-dimensional Poisson equation neglects contributions from the other two spatial dimensions. Likely, the most important contribution is in the direction of Galactocentric radius $R$, which gives a matter density correction equal to
\begin{equation}
    \Delta \rho = \frac{1}{4\pi G R}\frac{\partial}{\partial R}
    \left[ R \frac{\partial\Phi(R,z)}{\partial R} \right],
\end{equation}
where the quantity in the large square brackets is equal to the square of the Galactic circular velocity ($v_c^2$). Because the circular velocity curve is close to flat in the solar neighbourhood, this correction is likely small. According to recent studies, the local slope of the circular velocity is roughly $\partial v_c / \partial R = -1.5~\kmsec\, \kpc^{-1}$ ($-1.7\pm 0.1~\kmsec\, \kpc^{-1}$ in \citealt{2019ApJ...871..120E}; $-1.33\pm 0.1~\kmsec\, \kpc^{-1}$ in \citealt{2020ApJ...895L..12A}), giving a negligible correction of $\Delta \rho \simeq -0.0016~\Msunppcc$.

The approximation of separability also neglects some corrections, which in Jeans theory are known as the `axial' and `tilt' terms, which have been measured to be small this close to the Galactic plane \citep{budenbender,Sivertsson:2017rkp}. As discussed by \cite{BT2008} and \cite{Zhang_2013}, and also tested on simulations by \cite{Garbari2011}, the assumption of separability for the gravitational potential should be valid at least to heights comparable to the disk's scale height ($|z| \lesssim 300~\pc$).

There can potentially be smaller scale fluctuations in the potential due to substructures in the Galactic disk. If such fluctuations do exist and are large enough to produce a significant bias, this could cause a discrepancy in the results of the stellar samples. Identifying and quantifying such spatially dependent systematic effects was the main reason for dividing the local neighbourhood into sub-regions.

\subsection{Phase-space density}
In our model of inference we performed a joint fit of the gravitational potential and the full three-dimensional velocity distribution. Although our stellar tracer phase-space distribution was modeled to vary only as a function of distance from the Galactic mid-plane, the full three-dimensional velocity distribution was fitted for the following reason. While the majority of stars included in our stellar samples have high quality data which constrains their three-dimensional velocity very well, a significant subset of stars have missing radial velocity information and/or high astrometric uncertainties. The available information of such stars is still informative and helps constrain the population model, but their missing components need to be marginalised over. For example, for a star with missing radial velocity information, this velocity component is constrained by the overall population of stars; at the same time, the star's more precise data components are used to constrain the other two directions of the velocity distribution.

For the velocity distribution we assumed mirror symmetry across the Galactic plane (i.e. invariance with respect to $w \rightarrow -w$). For the directions parallel to the plane ($U$ and $V$), no such symmetry was imposed. In this sense, our model assumed that the phase-space distribution is well phase-mixed in the vertical direction, but not in the directions parallel to the Galactic plane. This is to be expected because the vertical oscillation frequency is comparatively high, especially for stars with low vertical energies, and therefore associated with a short relaxation time. Observations also indicate better phase-mixing in the vertical direction, whereas the $U$-$V$ plane is more structured (see for example figure 22 in \citealt{gaia_kinematics}). The $U$-$V$ distribution is largely distorted by the presence of several moving groups, many of which could have origin in resonances between the stars and the non-axisymmetric components of the Galaxy, like the bar and the spiral arms (see for example \citealt{2019A&A...626A..41M}).

We modeled the three-dimensional velocity distribution as a Gaussian mixture model, parametrised by Gaussian weights $a_k$, mean values $\bar{U}_k$ and $\bar{V}_k$ (due to mirror symmetry across the plane, $\bar{W}_k$ is fixed), and dispersions $\sigma_{U,k}$, $\sigma_{V,k}$, and $\sigma_{W,k}$. The index $k$ is only ever used to label the respective Gaussians of this mixture model. The velocity distribution in the Galactic mid-plane is proportional to
\begin{equation}
\begin{split}
    & f(z=0~\pc, \boldsymbol{V}) ~ \de^3\boldsymbol{V}
    \propto \\
    & \sum_k a_k\, \mathcal{M} \Bigg( \,
    \begin{bmatrix}
	U-\bar{U}_k \\
    V-\bar{V}_k \\
    W+W_\odot \\
   \end{bmatrix}, \,
    \begin{bmatrix}
	\sigma_{U,k}^2 & 0 & 0 \\
    0 & \sigma_{V,k}^2 & 0 \\
    0 & 0 &\sigma_{W,k}^2 \\
   \end{bmatrix} \, \Bigg) ~ \de^3\boldsymbol{V},
\end{split}
\end{equation}
where $\mathcal{M}$ represents the multivariate Gaussian distribution, defined
\begin{equation}\label{eq:multivariate_Gaussian}
	\mathcal{M}(\boldsymbol{p},\boldsymbol{\Sigma}_{\boldsymbol{p}}) \equiv
    \frac{\exp\left(-\dfrac{1}{2} \boldsymbol{p}^\top\boldsymbol{\Sigma}_{\boldsymbol{p}}^{-1}\boldsymbol{p} \right)}{\sqrt{(2\pi)^q \, | \boldsymbol{\Sigma}_{\boldsymbol{p}} |}},
\end{equation}
where $q$ is the number of dimensions of the column vector $\boldsymbol{p}$, and $\boldsymbol{\Sigma}_{\boldsymbol{p}}$ is a covariance matrix.

Because the phase-space distribution was assumed to be separable in the vertical direction and in a steady state configuration, the phase-space density at any height $z$ has the same distribution of vertical energy:
\begin{equation}
    E_z = \Phi(z)+w^2/2.
\end{equation}
Thus the phase-space density at any height $z$ is related to that of the mid-plane according to
\begin{equation}\label{eq:phase_space_density}
\begin{split}
    & f(z, \boldsymbol{V})\,\de^3\boldsymbol{V}
    \propto \\
    & \sum_k a_k\, \mathcal{M} \Bigg( \,
    \begin{bmatrix}
	U-\bar{U}_k \\
    V-\bar{V}_k \\
    \sqrt{w^2 + 2\Phi(z)} \\
   \end{bmatrix}, \,
    \begin{bmatrix}
	\sigma_{U,k}^2 & 0 & 0 \\
    0 & \sigma_{V,k}^2 & 0 \\
    0 & 0 &\sigma_{W,k}^2 \\
   \end{bmatrix} \, \Bigg)
   \,\de^3\boldsymbol{V},
\end{split}
\end{equation}
where it is implicit that $w=W+W_\odot$ and $z=Z+Z_\odot$. The peculiar velocities of the Sun in the directions parallel to the Galactic plane ($U_\odot$ and $V_\odot$) can be ignored, as the phase-space density is assumed to be invariant in these directions.

If we integrate the phase-space density over the velocities, we get the stellar number density in units of inverse volume, which is proportional to
\begin{equation}\label{eq:nuofz}
 	n(z) \propto \sum_k a_k \exp\left[-\frac{\Phi(z)}{\sigma_{W,k}^2}\right].
\end{equation}

\subsection{Expected density and gravitational potential}\label{sec:expectation}
In this work, our approach was to produce results that are data-driven and as model independent as possible. For this reason, the gravitational potential in our model of inference did not rely on any a priori expectations for the baryonic and halo dark matter density profiles. However, we do compare our end result with the expected gravitational potential, which is presented here.

The baryonic matter density profile comes from \cite{Schutz:2017tfp}, who compiled a table of baryonic components using results from \cite{Flynn:2006tm}, \cite{2015ApJ...814...13M}, and \cite{0004-637X-829-2-126}. This density profile has also been used in for example \cite{Sivertsson:2017rkp}, \cite{Buch:2018qdr}, and \cite{Widmark2019}.

\begin{figure}
	\centering
	\includegraphics[width=\columnwidth]{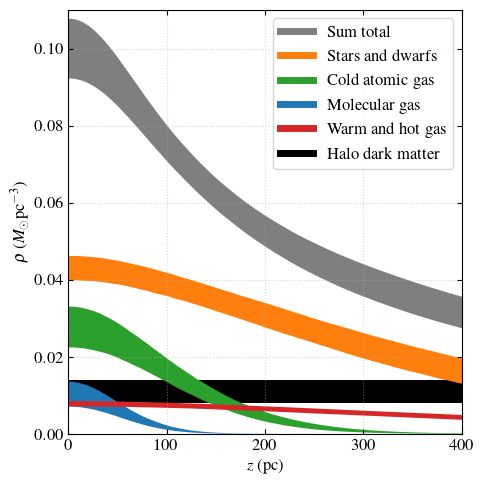}
    \caption{Expected total matter density shown as a $1\sigma$ band, also broken down into components of stars, cold atomic gas, molecular gas, hot gas, and halo dark matter.}
    \label{fig:expected_density}
\end{figure}

The baryonic components are listed in Table~\ref{tab:baryonic_components}. They are presented in terms of their mid-plane densities (a total of $0.0889\pm 0.0071~\Msunppcc$) and their respective velocity dispersions. Their respective vertical velocity distributions were assumed to be Gaussian and in equilibrium (i.e. iso-thermal), such that their matter densities (labelled by index $t$) varies with $z$ according to
\begin{equation}
    \rho_b = \sum_{t}
    \rho_{t}\exp\left[ -\frac{\Phi(z)}{\sigma_{t}^2} \right].
\end{equation}
In addition to the baryonic components, we also added a homogeneous dark matter halo density of $0.011\pm0.003~\Msunppcc$, with a conservatively large uncertainty \citep{Read2014}.\footnote{The exact value of the local dark matter density is still under debate. Although most results are in agreement with a value of $0.011~\Msunppcc$, they do differ somewhat depending on the method, for example global Galactic mass modelling \citep{2020MNRAS.494.4291C}, rotation curve fitting \citep{2019JCAP...09..046K,2019JCAP...10..037D,2020arXiv200913523B} or vertical Jeans analysis of different regions and stellar populations \citep{Sivertsson:2017rkp,2020A&A...643A..75S}.} The density components and how they vary with $z$ are summarised in Fig.~\ref{fig:expected_density}.

An important property of the expected matter density is that the shape of the baryonic density profiles are extrapolated using the components' observed vertical velocity dispersions. The vertical velocity distribution is also assumed to be Gaussian, which is a dubious assumption and somewhat discrepant with respect to this work's observed population of stars. In Appendix~\ref{app:expectation_stars_comparison} we compare the expected stellar density profile for stars with those of the stellar samples used in this work. Even larger systematic issues are associated with the baryonic gas components of the Galactic disk. Cold gas is difficult to observe and depends on some quite uncertain quantities: measurements of molecular hydrogen depend on the CO-to-H$_2$ conversion factor and 21 cm based measurements of atomic hydrogen depend on corrections for optical depth (see for example \citealt{2015A&A...579A.123H} for an instructive discussion). The contribution of the gaseous components to the gravitational potential is further complicated by the gas' non-uniform spatial structure in density and ionisation fraction; for example, the very local volume within roughly $100~\pc$ has a very low density of gas and dust \citep{2003A&A...411..447L}. In summary, the baryonic model presented here suffers from some potentially quite severe shortcomings and the systematic error could well be significantly larger than the statistical uncertainty. In Sects.~\ref{sec:discussion} and \ref{sec:conclusion} we discuss the need to update the baryonic model using more recent survey data.

{\renewcommand{\arraystretch}{1.6}
\begin{table}
\caption{Mid-plane densities $\rho_{t}$ and vertical velocity dispersions $\sigma_{t}$ of the baryonic matter components.}
\label{tab:baryonic_components}      
\centering          
\begin{tabular}{l l l} 
\hline
Component & $\rho_{t}~(\Msunppcc)$ & $\sigma_{t}~(\kmsec)$ \\
\hline
Molecular gas & $0.0104\pm 0.00312$ & $3.7\pm 0.2$ \\
Cold atomic gas & $0.0277\pm 0.00554$ & $7.1\pm 0.5$ \\
Warm atomic gas & $0.0073\pm 0.0007$ & $22.1\pm 2.4$ \\
Hot ionised gas & $0.0005\pm 0.00003$ & $39.0\pm 4.0$ \\
Giant stars & $0.0006\pm 0.00006$ & $15.5\pm 1.6$ \\
Stars, $M_V<3$ & $0.0018\pm 0.00018$ & $7.5\pm 2.0$ \\
Stars, $3<M_V<4$ & $0.0018\pm 0.00018$ & $12.0\pm 2.4$ \\
Stars, $4<M_V<5$ & $0.0029\pm 0.00029$ & $18.0\pm 1.8$ \\
Stars, $5<M_V<8$ & $0.0072\pm 0.00072$ & $18.5\pm 1.9$ \\
Stars, $M_V>8$ & $0.0216\pm 0.0028$ & $18.5\pm 4.0$ \\
White dwarfs & $0.0056\pm 0.0010$ & $20.0\pm 5.0$ \\
Brown dwarfs & $0.0015\pm 0.0005$ & $20.0\pm 5.0$ \\
\hline                  
\end{tabular}
\end{table}}

\section{Data and sample construction}\label{sec:data}
In this work we used data from \emph{Gaia} DR2. The data of a single star is written $\data_i$, labelled by the index $i$, and is listed in Table~\ref{tab:parameters}. It consists of the following information: apparent magnitude, $\hat{m}_{G,i}$; Galactic longitude and latitude, $\hat{l}_i$ and $\hat{b}_i$; parallax, $\hat{\varpi}_i$; proper motions in the latitudinal and longitudinal directions, $\hat{\mu}_{l,i}$ and $\hat{\mu}_{b,i}$; error covariance matrix for parallax and proper motions, $\hat{\Sigma}_i$; radial velocity, $\hat{v}_{RV,i}$; and radial velocity uncertainty, $\hat{\sigma}_{RV,i}$. The radial velocity is available only for a subset of stars in \emph{Gaia} DR2. For $\hat{m}_{G,i}$, $\hat{l}_i$, and $\hat{b}_i$, the observational uncertainties are small and can safely be neglected.

{\renewcommand{\arraystretch}{1.6}
\begin{table*}[ht]
	\centering
	\caption{Population parameters, stellar parameters, and data of our Bayesian hierarchical model.}
	\label{tab:parameters}
    \begin{tabular}{| l | l |}
		\hline
		$\popp$  & Population parameters \\
		\hline
		$\rho_{h = \{40,80,160,320\}}$ & Mid-plane densities of matter components with different scale heights \\
		$a_k$ & Weights of the mid-plane velocity distribution (with the constraint $\sum a_k=1$) \\
		$\sigma_{U,k},\, \sigma_{V,k}, \, \sigma_{W,k}$ & Dispersions of the velocity distribution \\
		$\bar{U}_k,\, \bar{V}_k$ & Mean values of the velocity distribution \\
		$Z_\odot$ & Height of the Sun with respect to the Galactic plane \\
		$W_\odot$ & Vertical velocity of the Sun in the rest frame of the Galactic plane \\
        \hline
        \hline
        $\objp_{i=1,...,N}$  & Stellar parameters of the $i$th star \\
        \hline
        $\mathbf{X}_i = (X_i,Y_i,Z_i)$ & Spatial position with respect to the Sun  \\
        $\mathbf{V}_i = (U_i,V_i,W_i)$ & Velocity in the solar rest frame \\
        $M_{G,i}$ & Absolute magnitude in the \emph{Gaia} $G$-band \\
        \hline
        \hline
        $\data_{i=1,...,N}$ & Data of the $i$th star \\
        \hline
        $\hat{m}_{G,i}$ & Apparent magnitude in the \emph{Gaia} $G$-band \\
        $\hat{l}_i$, $\hat{b}_i$ & Galactic longitude and latitude \\
        $\hat{\varpi}_i$ & Parallax \\
        $\hat{\mu}_{l,i}$, $\hat{\mu}_{b,i}$  & Proper motions \\
        $\hat{\boldsymbol{\Sigma}}_i$ & Error covariance matrix for proper motions and parallax \\
        $\hat{v}_{RV,i}$, $\hat{\sigma}_{RV,i}$  & Radial velocity and associated uncertainty (not available for all stars) \\
        \hline
	\end{tabular}
\end{table*}}

We made no cuts on the quality of the data (such as removing stars with poor astrometric precision, for example). Rather, our model fully accounted for all significant observational uncertainties. For the parallax measurements, the zero-point offset of $-0.03~\mas$ was subtracted from the catalogue values \citep{2018A&A...616A...2L}.

In this paper, the data is always written with hats (for example $\hat{\varpi}_i$). A quantity written without a hat (for example $\varpi_i$) always refers to the true property of the star.

\subsection{Sample construction}
\label{sec:samplecuts}

\begin{figure}
	\centering
	\includegraphics[width=\columnwidth]{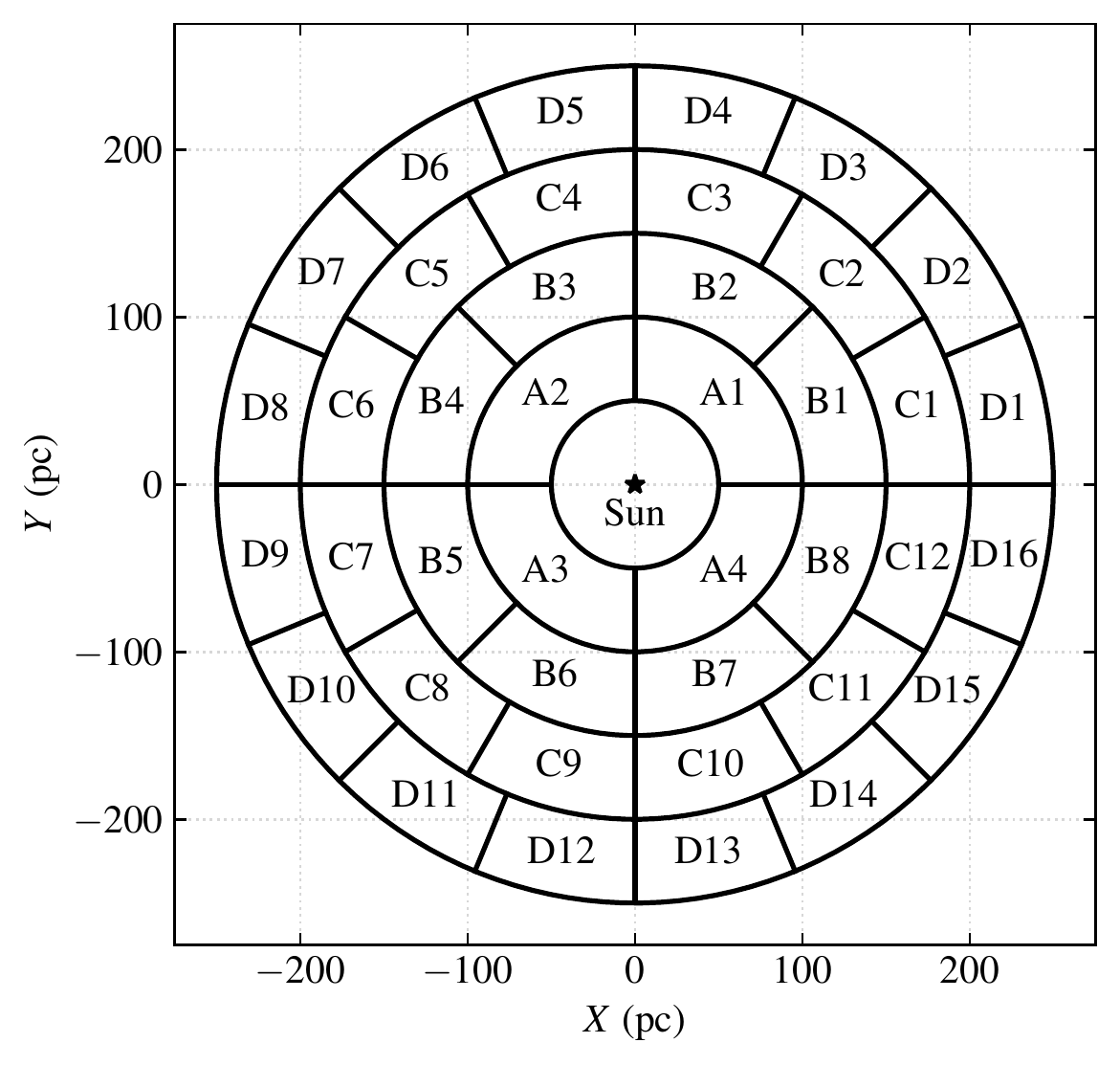}
    \caption{Area cells in the $X$-$Y$ plane by which we constructed our data samples. The point in the centre of the plot corresponds to the position of the Sun. For each area cell, we constructed three stellar samples reaching $\zlim = \{200, 300, 400\}~\pc$ above and below the Sun.}
    \label{fig:volumes}
\end{figure}

The stellar samples were constructed using cuts in spatial position and absolute magnitude. We made cuts in the spatial coordinates $X$ and $Y$ according to Fig.~\ref{fig:volumes}. This gave a total of 40 area cells, constructed to have roughly the same area in the $X$-$Y$ plane. They are labelled by a letter and a number, where the letter corresponds to the range in heliocentric cylindrical radius. The spatial volume of our stellar samples were limited in the vertical direction by a maximum height of $\zlim$ above and below the Sun, which took values of 200, 300, and 400 pc. This gave a total of 120 stellar samples (although samples that share area cell, and differ only in $\zlim$, are not statistically independent because they have a stellar membership overlap).

Each stellar sample has a shorthand name consisting of it's area cell followed by it's value of $\zlim$ in units of parsecs, for example \mbox{A1-200}. We also use the term region to refer to the area covered by all area cells beginning with the same letter, for example region A.

Because the true properties of the stars in the \emph{Gaia} DR2 catalogue are not perfectly known, the stellar sample had to be constructed using cuts on the data. Therefore, our volume cuts corresponded to the following criteria:
\begin{equation}\label{eq:volume_criteria}
\begin{split}
    \hat{l} & \in \Big[ (q-1) \frac{360^\circ}{4P},\, q \frac{360^\circ}{4P} \Big], \\
    \cos{\hat{b}}\,\frac{\mas}{\hat{\varpi}} & \in \Big[ P \times 0.05,\, (P+1) \times 0.05 \Big], \\
    \sin{\hat{b}}\,\frac{\mas}{\hat{\varpi}} & \in \Big[ -\frac{\zlim}{\kpc},\, \frac{\zlim}{\kpc} \Big], \\
\end{split}
\end{equation}
where $P = \{1,2,3,4\}$ for stellar sample names beginning with the letter $\{$A,B,C,D$\}$, and $q$ is the number in the sample name.

In order to obtain a homogeneous population of tracer stars, we also made cuts in absolute magnitude. The stellar sample should be homogeneous in the sense that the same cuts are applied to the stars' intrinsic properties, regardless of their phase-space coordinates. When constructing the stellar sample, we accounted for dust extinction using the three-dimensional maps of \cite{2017A&A...606A..65C}, and a conversion factor of 3.1 for going from $\text{E}(B-V)$ reddening to \emph{Gaia} $G$-band extinction \citep{dust_coeff,2018A&A...614A..19D}. The \emph{Gaia} $G$-band dust extinction as a function of spatial position is written $\mathcal{D}(l,b,\varpi)$. In terms of the data, the cut in absolute magnitude corresponded to the criteria
\begin{equation}\label{eq:absmag_criteria}
    \hat{m}_G-5\log_{10}\Bigg( \dfrac{\text{100~\mas}}{\hat{\varpi}} \Bigg) -\mathcal{D}(\hat{l},\hat{b},\hat{\varpi}) \in I_{M_G},
\end{equation}
where $I_{M_G}$ is the interval of allowed absolute magnitudes. Because the data quality of \emph{Gaia} DR2 is optimal in the range of apparent magnitudes of roughly 8--13 (mainly due to the availability of radial velocity information), the different area cells have slightly different absolute magnitude intervals, according to
\begin{itemize}
    \item region A: $I_{M_G} = [4.6, 5.5]$,
    \item region B: $I_{M_G} = [3.7, 5.0]$,
    \item region C: $I_{M_G} = [3.0, 4.7]$,
    \item region D: $I_{M_G} = [3.0, 4.5]$.
\end{itemize}

For each stellar sample, we also masked regions in $l$ and $b$ where there is an open cluster inside or in the vicinity of the sample volume, in order to avoid contamination from their peculiar spatial and kinematic properties. We used the catalogue of open clusters by \cite{OC_catalogue}. A mask was applied if the centre of the cluster was within the stellar sample volume or less than 50 pc from its spatial boundaries. Each masked region corresponded to a circular area on the sky, centred on the open cluster and with a radius of $5 \times r_{50}$, where $r_{50}$ is the angular radius that contains 50 \% of the open cluster's members. In terms of the data, the mask criteria can be stated like
\begin{equation}\label{eq:OCmask_criteria}
    (\hat{l}, \hat{b}) \notin \text{masked regions}.
\end{equation}
The percentage of the angular area that was masked by open clusters is listed in Table~\ref{tab:OCmasks}. A star in the \emph{Gaia} DR2 catalogue was included in the stellar sample if and only if all of the criteria in Eqs.~\eqref{eq:volume_criteria}, \eqref{eq:absmag_criteria}, and \eqref{eq:OCmask_criteria} were fulfilled.

\begin{table}
	\centering
	\caption{Percentage of angular area that was masked due to open clusters. Stellar samples not included in this table had masks that covered at most $1~\%$ of their angular area. The rows represent area cells (labelled to the left), and the columns represent $\zlim$ (labelled at the top).}
	\label{tab:OCmasks}
    \begin{tabular}{ l | r r r }
		 & $200~\pc$ & $300~\pc$ & $400~\pc$ \\
		\hline
		A2		& 1.3		& 1.3		& 1.3 \\
        A3		& 2.6		& 2.4		& 2.5 \\
        A4		& 4.8		& 4.6		& 4.5 \\
        B3		& 1.0		& 1.1		& 1.0 \\
        B4		& 13.7		& 13.0		& 12.8  \\
        B5		& 1.6		& 1.5 		& 1.5 \\
        B6		& 10.3		& 9.5		& 9.4 \\
        B7		& 12.5		& 11.7		& 11.6 \\
        C4		& 4.8 		& 4.4		& 4.1 \\
        C5		& 16.1		& 14.4		& 14.0 \\
        C6		& 8.6		& 7.8		& 7.3 \\
        C7		& 2.5		& 2.3		& 2.0 \\
        C9		& 11.6		& 10.5		& 10.1 \\
        C10		& 20.7		& 18.3		& 17.8 \\
        C12		& 8.4		& 7.5		& 7.2 \\
        D1		& 1.8		& 1.5		& 1.4 \\
        D4		& 1.1		& 0.8		& 0.8 \\
        D5		& 10.5		& 9.1		& 8.4   \\
        D6		& 2.7		& 2.3		& 2.1 \\
        D7		& 27.7		& 23.6		& 21.9 \\
        D10		& 5.6		& 4.7		& 4.4 \\
        D11		& 1.9		& 1.6		& 1.5 \\
        D12		& 23.6		& 20.1		& 18.6 \\
        D13		& 11.5		& 9.7		& 9.2 \\
        D14		& 3.2		& 2.7		& 2.6 \\
        D16		& 14.9		& 12.6		& 11.6 \\
	\end{tabular}
\end{table}

The spatial volume closest to the Sun ($\sqrt{X^2+Y^2}<50~\pc$) was ignored altogether in order to avoid systematic errors. A crucial requirement when constructing a sample of tracers is to select a population of stars that is uniform over the sample's spatial volume. Firstly, the completeness of \emph{Gaia} is poorly modeled for bright objects due to a lower number count, and bright objects are included by necessity in a volume that has no lower distance boundary. Secondly, multiple stellar systems, which are ubiquitous in the Milky Way \citep{Moe17}, can bias the observed number density of the tracer stars. A binary stellar system with a fixed orbital separation can be unresolved in outskirts of the sample volume, but resolved and seen as two stars in the Sun's immediate vicinity, breaking the requirement of homogeneity. For these reasons, it was safer to construct stellar samples with a small ratio between their most distant and nearest spatial points (a ratio which is infinite in the absence of a lower distance boundary).

The total number of stars in our stellar samples, for the different regions and values of $\zlim$, are presented in Table~\ref{tab:number_of_stars}. The stellar samples belonging to region D has a slightly lower number count, partly due to a lower stellar density for stars with brighter magnitudes. The outer regions also have a higher variance in the number counts, which is due to a higher rate of open cluster masks.

{\renewcommand{\arraystretch}{1.6}
\begin{table}
\caption{Number of stars in the stellar samples used in this work. They are presented in terms of their mean value and standard deviation, for groups determined by the different regions in the $x$-$y$ plane and different values of $\zlim$.}
\label{tab:number_of_stars}      
\centering          
\begin{tabular}{r | c c c}
Region & $\zlim = 200~\pc$ & $\zlim = 300~\pc$ & $\zlim = 400~\pc$ \\
\hline
A & $4783 \pm 218$ & $6274 \pm 202$ & $7335 \pm 176$ \\
B & $4713 \pm 375$ & $6225 \pm 432$ & $7315 \pm 454$ \\
C & $4722 \pm 449$ & $6195 \pm 498$ & $7256 \pm 517$ \\
D & $3806 \pm 512$ & $5001 \pm 577$ & $5849 \pm 604$ \\
\end{tabular}
\end{table}}

\section{Statistical model}\label{sec:statisticalmodel}
In this section we present our statistical model of inference. It is a Bayesian hierarchical model, which in this case has a hierarchy of two levels: the top level of population parameters $\popp$, which describe the population of tracer stars as a whole; and the bottom level of stellar parameters $\objp_i$, which describe the properties of individual stars.

The population parameters $\popp$ are listed in Table~\ref{tab:parameters}, and include: four parameters $\rho_{h = \{40,80,160,320\}}$, which determine the gravitational potential according to Eq.~\eqref{eq:gravitational_potential}; $a_k,\, \sigma_{U,k},\, \sigma_{V,k}, \, \sigma_{W,k},\, \bar{U}_k,\, \bar{V}_k$ are the weights, dispersions, and mean values of the three-dimensional velocity distribution, described in Eq.~\eqref{eq:phase_space_density}; $Z_\odot$ is the height of the Sun with respect to the Galactic plane; and $W_\odot$ is the vertical velocity of the Sun in the Galactic rest frame. The weights $a_k$ are constrained to sum to unity. The population parameters have $5+6K$ degrees of freedom, where $K$ is the number of Gaussians in the velocity distribution Gaussian mixture model.

The stellar parameters $\objp_i$, also listed in Table~\ref{tab:parameters}, are unique for each star and consist of: three-dimensional position, $\boldsymbol{X}_i$; three-dimensional velocity, $\boldsymbol{V}_i$; and absolute magnitude in the \emph{Gaia} $G$-band, $M_{G,i}$. The stellar parameters have a total of $7N$ degrees of freedom, where $N$ is the number of stars in the stellar sample.

The posterior density is the probability of $\popp$ and $\objp_i$ given $\data_i$, and is proportional to
\begin{equation}\label{eq:posterior_density}
\begin{split}
	& \text{Pr}(\popp, \, \objp_{i=\{1,...,N\}} | \, \data_{i=\{1,...,N\}} )\\
	& \propto \text{Pr}(\popp)
    \prod_{i=1}^N
    \frac{S(\data_i) \, \text{Pr}(\data_i \, | \, \objp_{i}) \, \text{Pr}(\objp_{i} \, | \, \popp)}{\bar{N}(\popp,S)},
\end{split}
\end{equation}
where $\text{Pr}(\popp)$ is the prior probability of the population parameters, $S(\data_i)$ is the selection function, $\text{Pr}(\data_i \, | \, \objp_{i})$ is the data likelihood, 
$\text{Pr}(\objp_{i} \, | \, \popp)$ is the stellar parameter probability density, and $\bar{N}(\popp,S)$ is a normalisation factor to the density distribution of stellar parameters. These five factors are described in detail in Sects.~\ref{sec:prior}--\ref{sec:normalisation} below. When sampling the posterior using Markov chain Monte Carlo (MCMC), only the relative fraction between different posterior values are relevant, such that any constant factors can be ignored. For this reason, many factors in the posterior density are written with a proportionality sign.

\begin{figure}
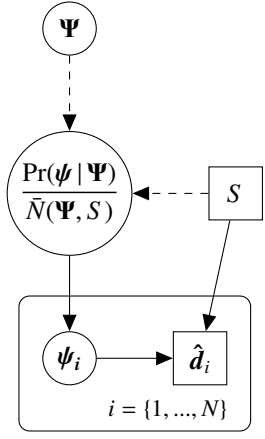

  \centering
  \tikz{ %
    \node[latent] (popp) {$\popp$} ; %
    \node[latent, below=of popp] (distfunc) {$\dfrac{\Pr(\objp \, | \, \popp)}{\bar{N}(\popp,S)}$} ; %
    \node[latent, right=of distfunc, rectangle] (selection) {$S$} ; %
    \node[latent, below=of distfunc] (objp) {$\boldsymbol{\objp_i}$} ; %
    \node[latent, right=of objp, rectangle] (data) {$\data_i$} ; %
    \plate[inner sep=0.3cm, xshift=-0.0cm, yshift=0.16cm] {plate} {(data) (objp)} {$i=\{1,...,N\}$}; %
    \edge[dashed] {popp} {distfunc} ; %
    \edge[dashed] {selection} {distfunc} ; %
    \edge {selection} {data} ; %
    \edge {distfunc} {objp} ; %
    \edge {objp} {data} ; %
  }
  \caption{Directed acyclic graph representing the Bayesian hierarchical model. Quantities inscribed in circles (squares) are free to vary (fixed) during the inference. Arrows with dashed (solid) lines represent deterministic (probabilistic) dependencies. The large square with rounded corners represents iteration over the stars in the stellar sample.}
  \label{fig:DAG}
\end{figure}

The statistical model is also represented in Fig.~\ref{fig:DAG} as a directed acyclic graph, illustrating how the top level population parameters set the probability distribution by which the bottom level stellar parameters are generated, which in turn generate the data. Because both $\popp$ and $\objp_i$ are free parameters, this model fully accounted for the uncertainties associated with the observation of each star.

The statistical model has a high number of free parameters (typically several $10^4$); to make it computationally tractable, it was implemented in \textsc{TensorFlow} which allows for auto-differentiation and posterior sampling using an MCMC algorithm called Hamiltonian Monte-Carlo. Further details on this can be found in Appendix~\ref{app:sampling}.

\subsection{Prior}\label{sec:prior}
The prior over the population parameters $\Pr(\popp)$ is a uniform box, although with the additional constraint that
\begin{equation}
    \sum_k a_k = 1.
\end{equation}
The bounds for the parameters are the following:
\begin{itemize}
    \item $\rho_h \in [0,\, 0.3]~\Msunppcc$,
    \item $a_k \in [0,\, 1]$,
    \item $\sigma_{U,k},\, \sigma_{V,k},\, \sigma_{W,k} \in [0,\, 250]~\kmsec$,
    \item $\bar{U}_k,\, \bar{V}_k \in [-250,\, 250]~\kmsec$,
    \item $Z_\odot \in [-50, 50]~\pc$,
    \item $W_\odot \in [-20, 20]~\kmsec$.
\end{itemize}
This uniform box prior is wide enough for the posterior densities not to be affected by the specific position of its boundaries (with the exception of some bounds which prevent negative values).

\subsection{Selection function}
The selection function models the probability that a star will be included in the stellar sample. It is a function of data, written
\begin{equation}
    S(\data) = \mathcal{K}(\hat{l}, \hat{b}, \hat{\varpi}, \hat{m}_G) \times \mathcal{C}(\hat{l}, \hat{b}, \hat{m}_G).
\end{equation}
The factor $\mathcal{K}(\hat{l}, \hat{b}, \hat{\varpi}, \hat{m}_G)$ corresponds to the data cuts by which the stellar samples are constructed; it includes the cuts in $\hat{l}$, $\hat{b}$, $\hat{\varpi}$, and $\hat{m}_G$ that define the stellar sample, and also the open cluster masked regions in $\hat{l}$ and $\hat{b}$, as described in Sect.~\ref{sec:samplecuts}. The function $\mathcal{K}(\hat{l}, \hat{b}, \hat{\varpi}, \hat{m}_G)$ takes a value of one when all criteria are fulfilled, and is otherwise zero. The factor $\mathcal{C}(\hat{l}, \hat{b}, \hat{m}_G)$ is the completeness of \emph{Gaia} DR2; it is a function of angular position and apparent magnitude, calculated using a cross-match with the 2MASS catalogue \citep{2018asclsoft11018R}. The average completeness in the relevant range of apparent magnitudes is evaluated to around $99~\%$.

\subsection{Likelihood}
The likelihood of the data, given a set of stellar parameters (here parametrised in the space of observables and denoted by quantities without hats), is proportional to
\begin{equation}\label{eq:likelihood}
\begin{split}
	& \text{Pr}(\data_i \, | \objp_i) \propto
	\delta(l_i-\hat{l}_i) \times \delta(b_i-\hat{b}_i) \times \delta(m_{G,i}-\hat{m}_{G,i}) \times    \\
	& \mathcal{M}\left(
    \begin{bmatrix}
    	\mu_{l,i} - \hat{\mu}_{l,i} \\
        \mu_{b,i} - \hat{\mu}_{b,i} \\
        \varpi_i - \hat{\varpi}_i \\
    \end{bmatrix}, \hat{\boldsymbol{\Sigma}}_i
    \right)\,
    \times \mathcal{N}(v_{RV,i}-\hat{v}_{RV,i},\hat{\sigma}_{RV,i}),
\end{split}
\end{equation}
where $\delta$ is the Dirac delta function (uncertainties for $\hat{l}$, $\hat{b}$, and $\hat{m}_G$ are neglected), and
\begin{equation}
    \mathcal{N}(x, \sigma) \equiv 
    \dfrac{ \exp\Bigg( -\dfrac{x^2}{2\sigma^2} \Bigg) }{\sqrt{2\pi\sigma^2}}
\end{equation}
is the one-dimensional Gaussian distribution. The error covariance matrix
\begin{equation}
	\hat{\boldsymbol{\Sigma}}_i = 
    \begin{bmatrix}
	\hat{\sigma}_{\mu_l}^2 & \hat{\rho}_{\mu_l\mu_b}\hat{\sigma}_{\mu_l}\hat{\sigma}_{\mu_b} & \hat{\rho}_{\mu_l\varpi}\hat{\sigma}_{\mu_l}\hat{\sigma}_{\varpi} \\
    \hat{\rho}_{\mu_l\mu_b}\hat{\sigma}_{\mu_l}\hat{\sigma}_{\mu_b} & \hat{\sigma}_{\mu_b}^2 & \hat{\rho}_{\mu_b\varpi}\hat{\sigma}_{\mu_b}\hat{\sigma}_{\varpi} \\
    \hat{\rho}_{\mu_l\varpi}\hat{\sigma}_{\mu_l}\hat{\sigma}_{\varpi} & \hat{\rho}_{\mu_b\varpi}\hat{\sigma}_{\mu_b}\hat{\sigma}_{\varpi} & \hat{\sigma}_{\varpi}^2
   \end{bmatrix}
\end{equation}
accounts for all uncertainties $\hat{\sigma}$ and uncertainty correlations $\hat{\rho}$ between the measured parallax and proper motions of the $i$th star (for shorthand, the index $i$ is dropped in the right-hand side of the above expression). If radial velocity information is not available, the factor $\mathcal{N}(v_{RV,i}-\hat{v}_{RV,i},\hat{\sigma}_{RV,i})$ is dropped from Eq.~\eqref{eq:likelihood}.

\subsection{Stellar parameter probability density}
The stellar parameter probability density for a star, labelled by the index $i$, is equal to
\begin{equation}
    \text{Pr}(\objp_{i} \, | \, \popp) =
    \text{Pr}(\boldsymbol{X}_i,\boldsymbol{V}_i \, | \, \popp) \times
    \text{Pr}(M_{G,i}),
\end{equation}
where $\text{Pr}(\boldsymbol{X}_i,\boldsymbol{V}_i \, | \, \popp)$ the phase-space probability density, and $\text{Pr}(M_{G,i})$ is the probability of its absolute magnitude (which is independent of $\popp$).

For almost all stars in our stellar samples the distance is constrained well enough such that their inferred absolute magnitude varies only minimally. Hence, including $\text{Pr}(M_{G,i})$ in the posterior density grants no significant power of inference for such stars. It is relevant only for the few stars with very poor parallax precision, in order to suppress the probability that they are extremely bright objects very far away. For this reason we modeled the distribution of magnitudes in our stellar parameter probability density according to the following simple form,
\begin{equation}
    \text{Pr}(M_{G,i}) \propto \tanh\Bigg(\frac{M_{G,i}-3.701}{1.745}\Bigg) + 1,
\end{equation}
which was fitted to the distribution of solar neighbourhood stars with $M_G < 7$.

The phase-space probability is given by Eq.~\eqref{eq:phase_space_density}, according to
\begin{equation}
    \text{Pr}(\boldsymbol{X}_i,\boldsymbol{V}_i \, | \, \popp) ~
    \de^3\boldsymbol{X}_i \,
    \de^3\boldsymbol{V}_i =
    f(z_i,\boldsymbol{V}_i \, | \, \popp) ~ \de^3\boldsymbol{X}_i \,
    \de^3\boldsymbol{V}_i,
\end{equation}
where we now write $f(z_i,\boldsymbol{v}_i \, | \, \popp)$ with an explicit dependence on $\popp$. The differential factors are written in order to be explicit that the above expression is written in terms of the phase-space coordinates $\boldsymbol{X}$ and $\boldsymbol{V}$.

When we sampled our posterior probability density, we reparametrised the phase space coordinates $\boldsymbol{X}_i$ and $\boldsymbol{V}_i$ to the space of observables ($l_i$, $b_i$, $\varpi_i$, $\mu_{l,i}$, $\mu_{b,i}$, $v_{RV,i}$). Because uncertainties on $l$ and $b$ are neglected, those coordinates are fixed by the data. Combined with the angle factors of the data likelihood, the phase-space density can be reformulated like
\begin{equation}
\begin{split}
    & \delta(l_i-\hat{l}_i) \times \delta(b_i-\hat{b}_i) \times f(Z_i+Z_\odot,\boldsymbol{V}_i \, | \, \popp) ~ \de^3\boldsymbol{X}_i \,
    \de^3\boldsymbol{V}_i \propto \\
    & f[ Z_i(...)+Z_\odot,\boldsymbol{V}_i(...) \, | \, \popp ]\,
    \varpi_i^{-6}
    ~ \de\varpi_i \, \de\mu_{l,i} \, \de\mu_{b,i}\, \de v_{RV,i},
\end{split}
\end{equation}
where the additional factor $\varpi_i^{-6}$ comes from the Jacobian of the coordinate transformation (constant factors neglected). In this expression it is implicit that $Z_i(...)$ and $\boldsymbol{V}_i(...)$ depend on the observables and angular data, according to
\begin{equation}
    Z_i(...) = \sin(\hat{b}_i)\frac{\mas \times \kpc}{\varpi_i}
\end{equation}
and
\begin{equation}
    \boldsymbol{V}_i(...) = \boldsymbol{R}(\hat{l}_i, \hat{b}_i) \times
    \begin{bmatrix}
    k_\mu \times \mu_{l,i} / \varpi_i \\
    k_\mu \times \mu_{b,i} / \varpi_i \\
    v_{RV,i}
    \end{bmatrix},
\end{equation}
where $k_\mu = 4.74057 ~ \yr \times \kmsec$ and
\begin{equation}\label{eq:rotation_matrix}
	\boldsymbol{R}(l,b) =
    \begin{bmatrix}
    -\sin(l) &  -\cos(l)\sin(b) & \cos(l)\cos(b) \\
    \cos(l) & -\sin(l)\sin(b) & \sin(l)\cos(b)  \\
    0 & \cos(b) & \sin(b)
    \end{bmatrix}
\end{equation}
is a rotational matrix that transforms the longitudinal-latitudinal-radial directions to those of $\boldsymbol{V}$.

\subsection{Normalisation}\label{sec:normalisation}
Because there are two levels in our Bayesian hierarchical model, there are two levels of normalisation. The normalisation to the population parameters is fixed and therefore ignored, as any constant factor can be dropped from the posterior probability density. Conversely, the normalisation to the distribution of stellar parameters will vary, dependent on the population parameters. This normalisation factor is included in the denominator of Eq.~\eqref{eq:posterior_density}, written $\bar{N}(\popp,S)$. It ensures that a star, randomly generated from the population model, has an integrated probability of one to be included in the stellar sample. It is equal to
\begin{equation}\label{eq:norm}
	\bar{N}(\popp,S) =
    \int S(\data_x)\, \text{Pr}(\data_x \, | \, \objp_x)\, \text{Pr}(\objp_x \, | \, \popp) ~ \de\objp_x\, \de\data_x.
\end{equation}
The stellar parameters and data, $\objp_x$ and $\data_x$, have an index $x$ in order to highlight that this object is hypothetical and generated from the population model set by $\popp$.

The integral in Eq.~\eqref{eq:norm} is high dimensional and expensive to compute. However, because the phase-space probability density is invariant with respect to $X$ and $Y$, and because the selection function does not depend on velocity, most of this integral only needs to be computed once. The normalisation factor can be reformulated like
\begin{equation}\label{eq:norm_with_effective_area}
	\bar{N}(\popp,S) = \int n(z \, | \,  \popp)\,A_\text{eff}(Z \, | \, S)~\de Z,
\end{equation}
where $A_\text{eff}(Z \, | \, S)$ is an effective area which depends on the selection function $S$ and varies with $Z$. It is equal to
\begin{equation}\label{eq:effective_area}
\begin{split}
	& A_\text{eff}(Z \, | \, S) =
	\int S(\data_x)\, \text{Pr}(\data_x \, | \, \objp_x) ~ \de X\, \de Y\, \de\data_x \\
	& \propto
	\int S(\hat{l}, \hat{b}, \hat{\varpi}, \hat{m}_G)\, \text{Pr}(\hat{l}, \hat{b}, \hat{\varpi}, \hat{m}_G \, | \, X,Y,Z) ~ \de X\, \de Y\, \de \hat{l}\, \de \hat{b}\, \de \hat{\varpi}\, \de \hat{m}_G.
\end{split}
\end{equation}
In the second row of this expression, only the data components relevant for the selection function are integrated over. The effective area does not depend on the population parameters $\popp$, which means it can be computed separately and catalogued before computing the full posterior probability density.

The expression $\text{Pr}(\hat{l}, \hat{b}, \hat{\varpi}, \hat{m}_G \, | \, X,Y,Z)$ is the probability of generating a certain set of data given a star's true properties. The angles $\hat{l}$ and $\hat{b}$ are given directly by the spatial position. The apparent magnitude $\hat{m}_G$ is also given directly once we also generate the absolute magnitude, which is done from the observed distribution of absolute magnitudes in the solar neighbourhood, written $F(M_G)$ (this distribution is the same as used in \citealt{Widmark2019}, see section 2.4 in that paper for more details). Finally, the parallax $\hat{\varpi}$ has significant uncertainties, which are generated from the distribution of parallax uncertainties in the specific volume of the stellar sample. It is modeled as a function of apparent magnitude and latitude, written $h(\hat{\sigma}_\varpi\, |\, \hat{m}_G,\hat{b})$, given by a two-dimensional histogram with bin sizes of $\sim 0.2$ mag and $\sim 10$ degrees.

Thus we have that
\begin{equation}
\begin{split}
    & \text{Pr}(\hat{l}, \hat{b}, \hat{\varpi}, \hat{m}_G \, | \, X,Y,Z) =
    \delta[ l(X,Y,Z) - \hat{l} ] \times
    \delta[ b(X,Y,Z) - \hat{b} ] \\
    & \times \int \delta \Bigg[M_G + 5\log_{10}\Bigg( \dfrac{\text{100~\mas}}{\varpi(X,Y,Z)} \Bigg) -\hat{m}_G \Bigg] \, F(M_G) \, \de M_G \\
    & \times \int \mathcal{N}(\varpi-\hat{\varpi},\hat{\sigma}_\varpi)\,
    h(\hat{\sigma}_\varpi\, |\, \hat{m}_G,\hat{b})\,
    \de\hat{\sigma}_\varpi,
\end{split}
\end{equation}
where $\delta$ is the Dirac delta function.

In practice, we calculated $A_\text{eff}(Z \, | \, S)$ numerically by rejection sampling. This was done by randomly generating a star's position, given a uniform stellar number density, and the observational data according to the distribution of absolute magnitudes and parallax uncertainties described above. We took care to randomise stars whose true properties lied outside the region of the stellar sample, as stars can be scattered into the sample region through observational errors.

\begin{figure}
	\centering
	\includegraphics[width=0.8\columnwidth]{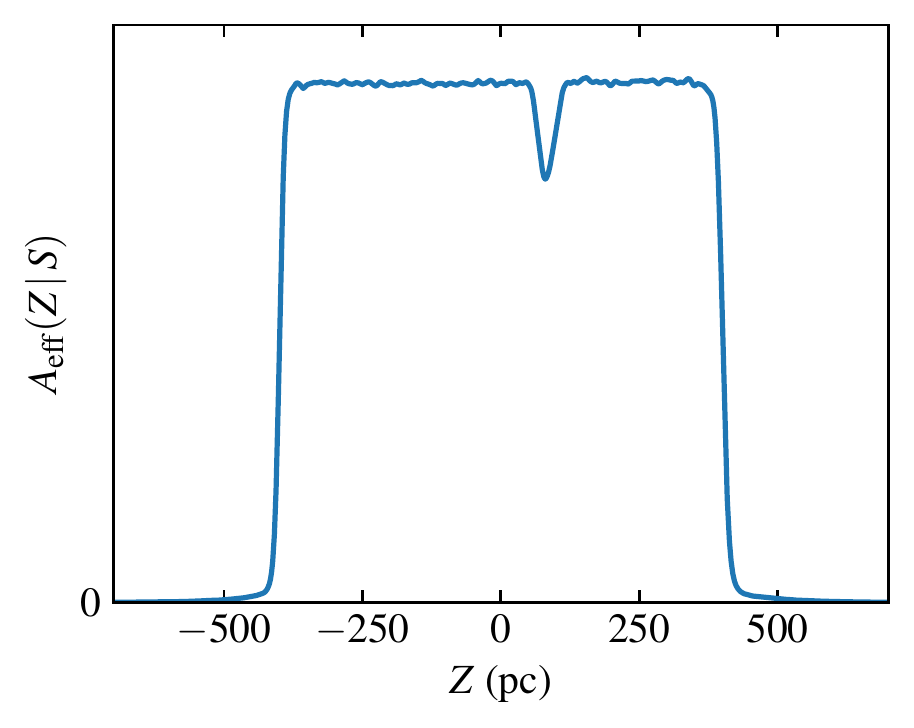}
    \caption{Effective area of stellar sample B5-400. The depression around $Z=80~\pc$ is due to an open cluster mask.}
    \label{fig:effective_area}
\end{figure}

An example of a stellar sample's effective area is shown in Fig.~\ref{fig:effective_area}. It differs from the idealised flat box of the stellar sample's geometric volume due to the completeness of \emph{Gaia} DR2, data uncertainties, and open cluster masks.

\section{Results}\label{sec:results}
In this section we discuss our results, mainly in terms of the inferred posterior probability densities. How the posterior density was sampled is discussed in more detail in Appendix~\ref{app:sampling}. For the velocity distribution we used a Gaussian mixture model consisting of 10 Gaussians, giving a total of 65 degrees of freedom for the population model. This number of Gaussians gave enough flexibility to produce good phase-space distribution fits. We also ran our inference using 20 Gaussians but the results did not change notably.

The results in this work are mainly presented in terms of the gravitational potential rather than the inferred matter density profiles. The gravitational potential is arguably the quantity that was actually being measured from which the matter density could be calculated via the Poisson equation (see Eq.~\ref{eq:Poisson}). Because the Poisson equation contains a second derivative, the matter density was not as robustly inferred; it is more dependent on the choice of parametrisation and priors, and suffers from strong degeneracies, especially between its mid-plane amplitude and profile shape. For this reason, the gravitational potential is the more illustrative quantity, and easier to compare with the expected result.

\begin{figure}
	\centering
	\includegraphics[width=\columnwidth]{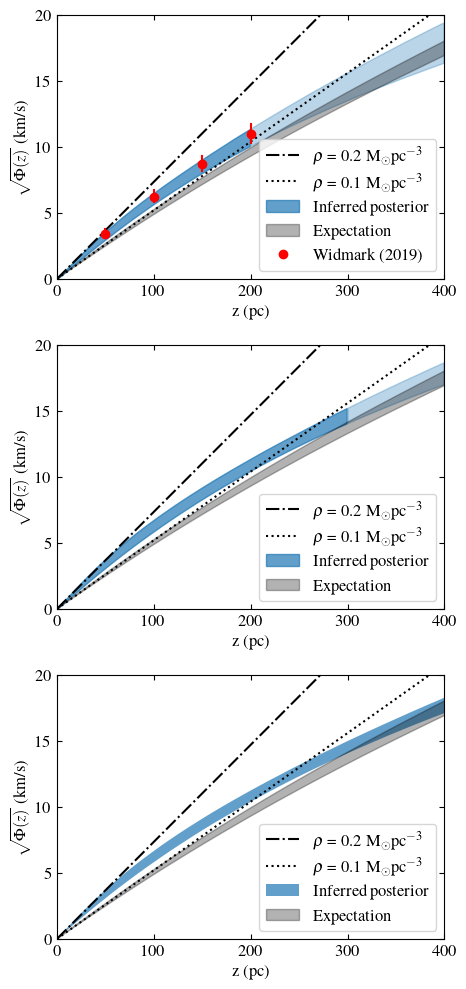}
    \caption{Inferred gravitational potential of stellar samples A1-200, A1-300, and A1-400, in the top, middle, and bottom panel, plotted in terms of its square root as a function of $z$. The width of the blue band shows the 16th to 84th percentile of the inferred posterior density of the respective stellar samples. The band has a lighter colour for $z>\zlim$, indicating that this is an extrapolation outside the spatial volume of the stellar sample. The $1\sigma$ band of the expected gravitational potential is shown in grey (identical in all panels). The black dash-dotted and dotted lines show the gravitational potential in case of a homogeneous matter density of 0.2 and 0.1 $\Msunppcc$. In the top panel, we also plot the inferred potential of \cite{Widmark2019} in red, where the dot indicates the median and the error bars cover the 10th and 90th percentiles of that work's eight stellar sample posterior densities.}
    \label{fig:A1_phis}
\end{figure}

The inferred gravitational potentials of stellar samples A1-200, A1-300, and A1-400 are shown in Fig.~\ref{fig:A1_phis}, and the three samples agree well with each other. They are consistent with the expected potential (see Sect.~\ref{sec:expectation}) at larger distances from the mid-plane, around $|z|\simeq 400~\pc$. However, the potential is significantly steeper at small distances of $|z| \lesssim 60~\pc$, implying a high mid-plane matter density that quickly decreases with $z$. In the top panel of Fig.~\ref{fig:A1_phis}, we also show the inferred potential of \cite{Widmark2019}. That work used different cuts in spatial volume and absolute magnitude\footnote{In \cite{Widmark2019}, all stellar samples were constructed from a spatial volume consisting of a spherical shell defined by a heliocentric distance between 100 and 200 pc. The eight samples had non-overlapping cuts in absolute magnitude, with $M_G$ ranging from 3 to 6.3.}, but the results agree well with this work.

\begin{figure}
	\centering
	\includegraphics[width=\columnwidth]{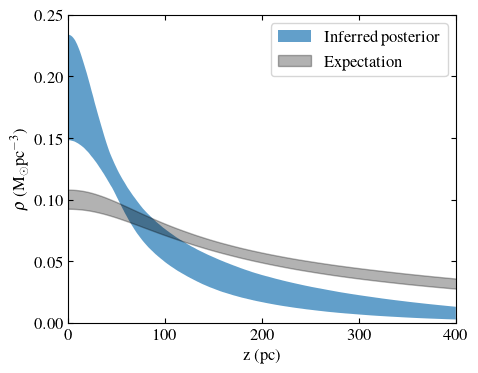}
    \caption{Inferred total matter density profile of stellar sample A1-400. The width of the blue band shows the 16th to 84th percentile of the inferred posterior density. The $1\sigma$ band of the expected gravitational potential is shown in grey.}
    \label{fig:A1_rho}
\end{figure}

The inferred total matter density as a function of $z$, for stellar sample A1-400, is shown in Fig.~\ref{fig:A1_rho}. The inferred matter density is very strongly concentrated to the Galactic mid-plane and decays quickly with height. As such, it does not imply a total matter density surplus if we consider the total surface density within a sufficiently large distance from the mid-plane. The inferred distribution is at odds with the expected matter density, as described in Sect.~\ref{sec:expectation}, which does not decay as quickly with height $|z|$; hence, this result cannot be reconciled, at least not completely, by including some kind of hidden over-density. If we use the expected gravitational potential, while keeping other population parameters fixed, we obtain a normalised stellar tracer density, $n(z)/\bar{N}(\popp,S)$, that is $\sim 5~\%$ higher in the mid-plane and $\sim 5~\%$ lower at $|z| \simeq 200~\pc$. This demonstrates that our result can be explained by a comparatively small time-varying feature in the stellar tracer density, rather than the dramatic difference in the total matter density seen in Fig.~\ref{fig:A1_rho}. See Sect.~\ref{sec:discussion} for further discussion on the likelihood that the assumption of a steady state is broken. Similar matter density profiles are obtained for practically all stellar samples.

\begin{figure*}
	\centering
	\includegraphics[width=\textwidth]{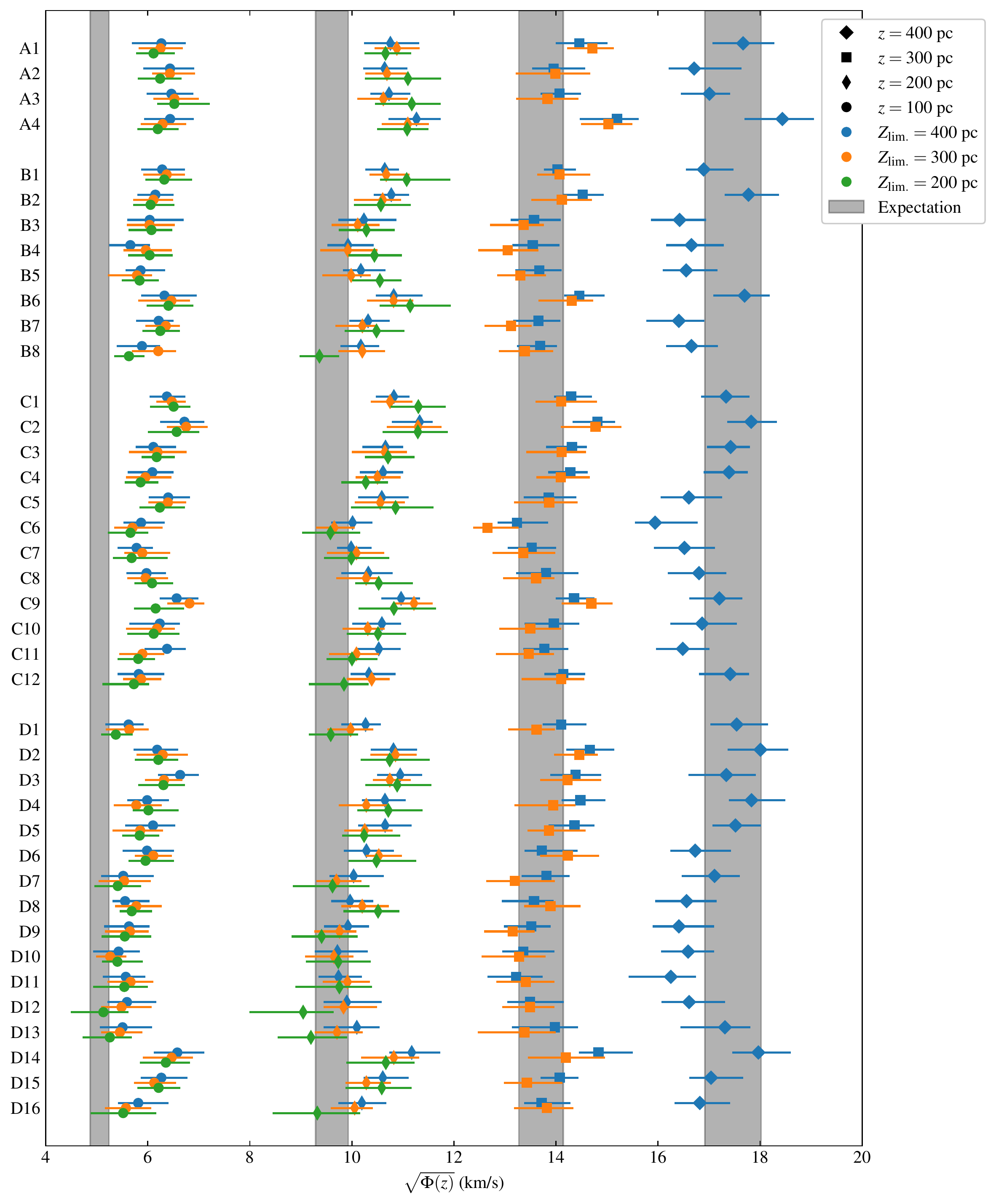}
    \caption{Inferred potential for the 120 stellar samples, presented in terms of $\sqrt{\Phi(z)}$, where $z = \{100,\, 200,\, 300,\, 400\}~\pc$. The samples are listed vertically in terms of area cells, and in subgroups of descending $\zlim$ (also labelled by colour according to the legend). Each respective marker shows the median gravitational potential of the inferred posterior distribution, and the marker's shape represents the height $z$ according to the legend. The horizontal lines represent the posterior widths, whose endpoints are equal to the 16th and 84th percentile of the posterior distribution. The $1\sigma$ bands of the expected gravitational potential at the respective heights are shown in grey.}
    \label{fig:phi_summary}
\end{figure*}

\begin{figure}
	\centering
	\includegraphics[width=\columnwidth]{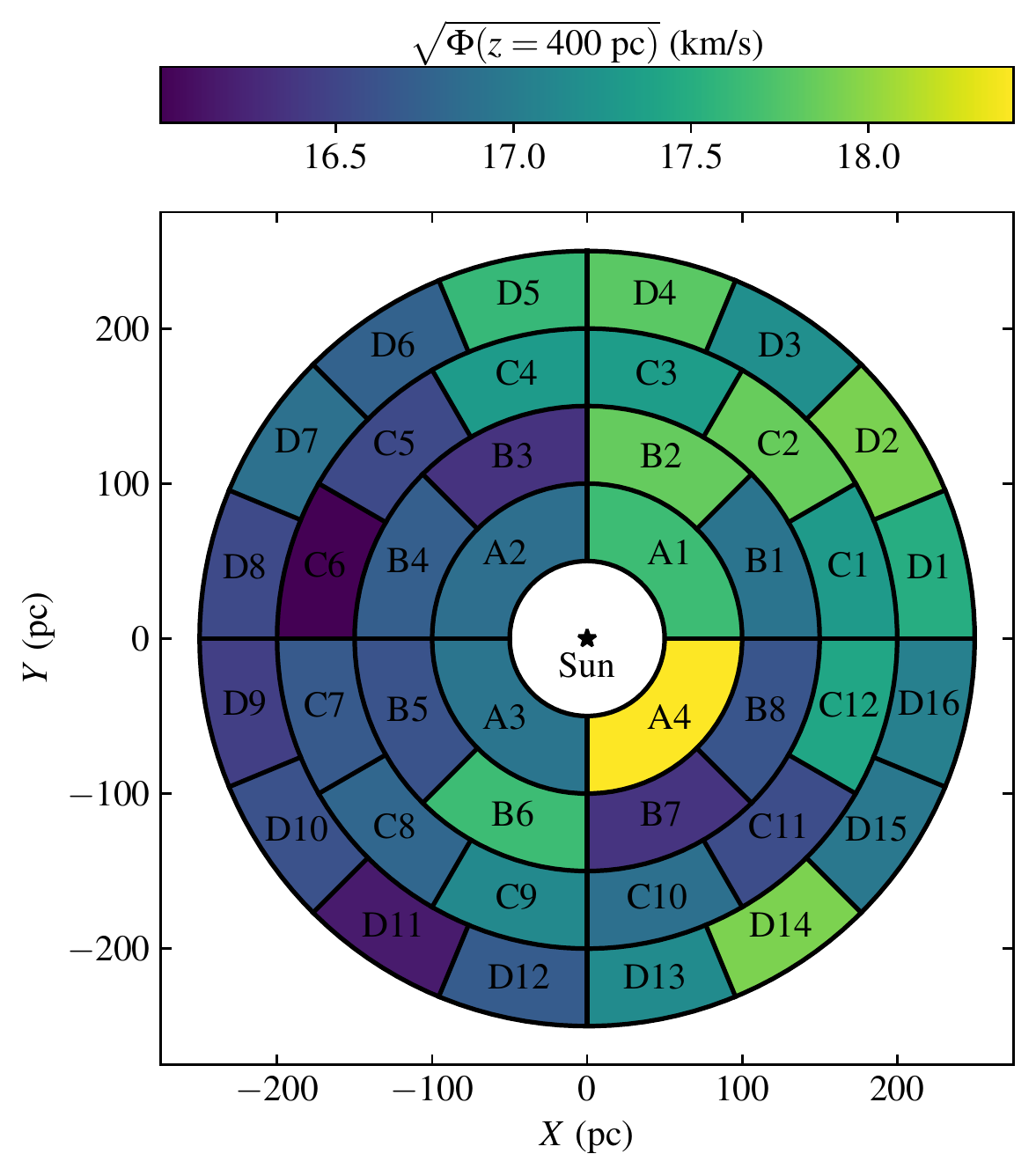}
    \caption{Inferred gravitational potential at $z=400~\pc$ for different area cells, presented in terms of the posterior distribution median, using stellar samples with $\zlim = 400~\pc$.}
    \label{fig:volumes_inferred_phis}
\end{figure}

In Fig.~\ref{fig:phi_summary}, we show the inferred potential of our 120 stellar samples, for heights $z=\{100,200,300,400\}~\pc$. For almost all stellar samples, we see the same general behaviour as for area cell A1, with a steep potential close to the Galactic mid-plane, but consistent with the expected potential at greater $|z|$. There are discrepancies between the inferred potential of the stellar samples, where B8-200 stands out with low values, and to a lesser extent also area cells D7, D9, D12, D13, and D16. The spatial dependence of the inferred potentials is illustrated in Fig.~\ref{fig:volumes_inferred_phis}, where the area cells are colour coded according to the median posterior value of $\Phi(z=400~\pc)$, for samples with $\zlim = 400~\pc$. Our model for the expected potential is evaluated at the Sun's position, but our stellar samples do span a significant range in the Galactic radius (i.e. $X$), over which the total matter density is expected to vary. We fitted the variation of $\Phi(z=400~\pc)$ with respect to $X$ for the stellar samples of region C, accounting for the statistical variance of the stellar samples' posterior densities, and obtain a disk scale length of a few kiloparsecs; the maximum likelihood is for 3.7 kpc, but this is associated with a large uncertainty, allowing 2.6 kpc within $1\sigma$ (similar results are found for region D, for which 2.6 kpc is consistent within $1.5\sigma$). This is consistent with recent studies: see for example the review by \cite{2016ARA&A..54..529B}, that report scale lengths for the thin and thick disks of $\sim 2.6~\kpc$ and $\sim 2\kpc$, where the former is dominant in our case.\footnote{The scale length of the Galactic disk also varies depending on the stellar population. For example, \cite{2012ApJ...753..148B} show a clear dependence on metallicity, where the most metal rich stars have scale lengths up to $\sim 4~\kpc$.} We do not see a statistically significant variation in the azimuthal direction (i.e. $Y$).

\begin{figure}
	\centering
	\includegraphics[width=\columnwidth]{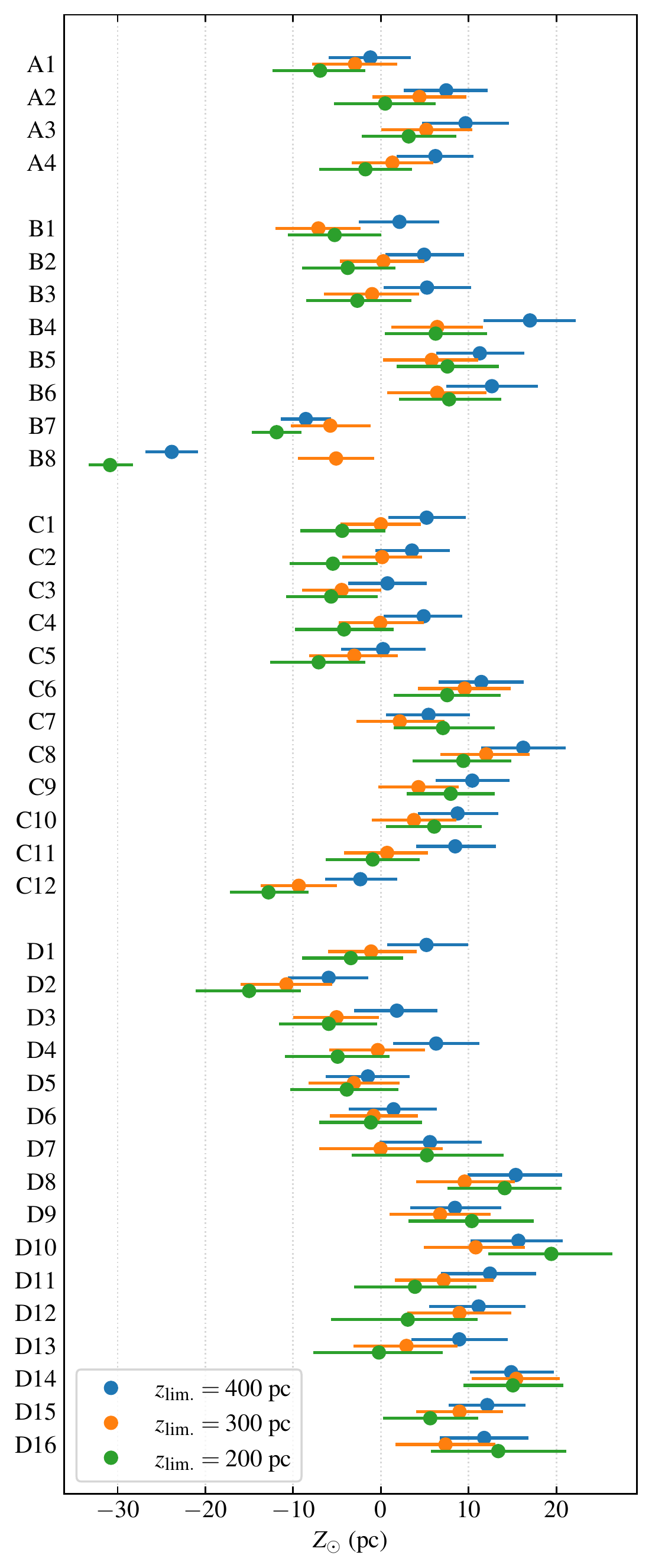}
    \caption{Inferred height of the Sun, $Z_\odot$, for the 120 stellar samples. The samples are listed vertically according to area cell, in groups of three for different $\zlim$ in descending order, which are labelled by colour according to the legend. The horizontal lines represent the posterior widths, whose endpoints are equal to the 16th and 84th percentile of the posterior distribution.}
    \label{fig:zsun_summary}
\end{figure}

\begin{figure}
	\centering
	\includegraphics[width=\columnwidth]{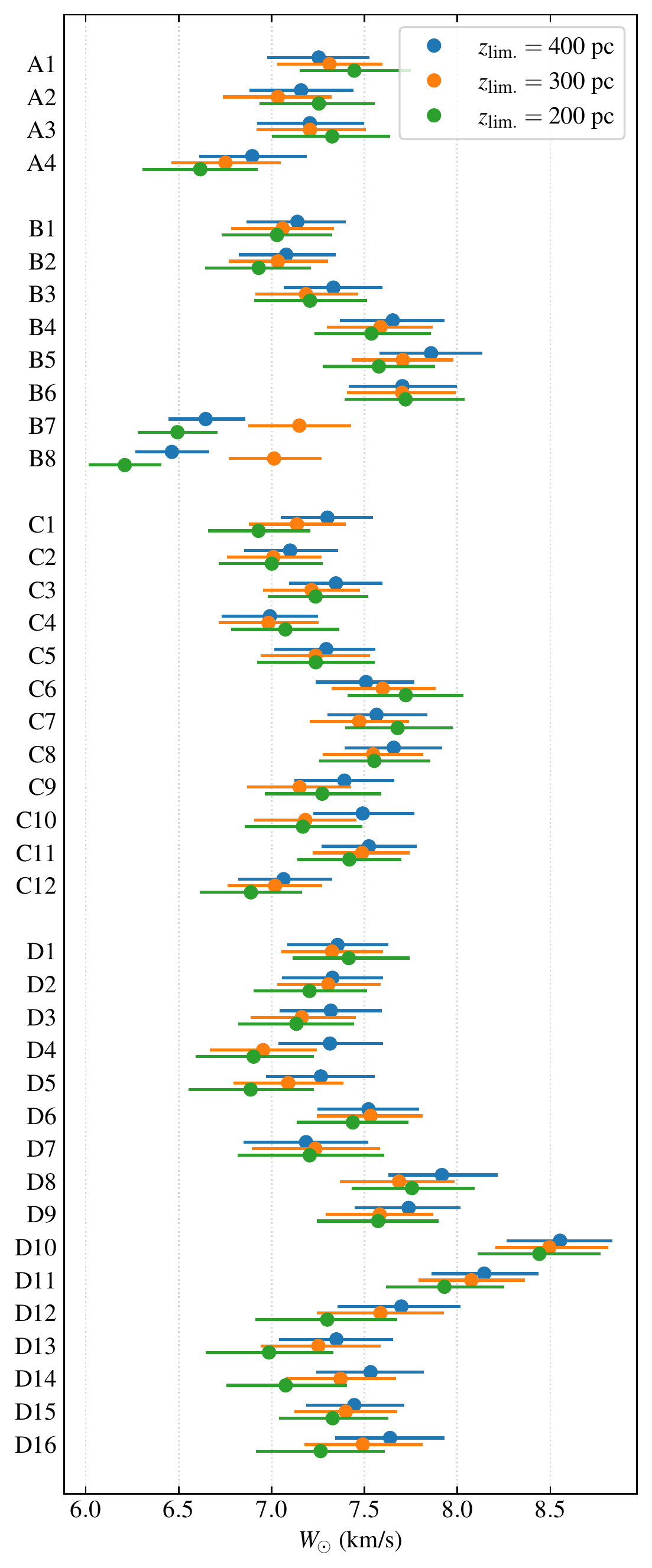}
    \caption{Inferred vertical velocity of the Sun, $W_\odot$, for the 120 stellar samples. The samples are listed vertically according to area cell, in groups of three for different $\zlim$ in descending order, which are labelled by colour according to the legend. The horizontal lines represent the posterior widths, whose endpoints are equal to the 16th and 84th percentile of the posterior distribution.}
    \label{fig:wsun_summary}
\end{figure}

In Figs.~\ref{fig:zsun_summary} and \ref{fig:wsun_summary} we show the inferred values for $Z_\odot$ and $W_\odot$ of our 120 stellar samples. There are discrepancies between stellar samples for both $Z_\odot$ and $W_\odot$, larger than statistical errors can account for, as well as a spatial dependence. For $Z_\odot$, the mean values of all stellar samples is $2.8~\pc$, and $\{0.4,2.0,6.0\}~\pc$ when grouping them by $\zlim = \{200,300,400\}~\pc$, which is indicative of a broken Galactic plane mirror symmetry. Furthermore, these values are low with respect to some other studies that reach kiloparsec distances from the Galactic plane ($25\pm 5~\pc$, $13.4\pm 4.4~\pc$, and $20.8\pm 0.3~\pc$, for \citealt{Juric:2005zr}, \citealt{2017MNRAS.468.3289Y}, and \citealt{BovyAssym}). For $W_\odot$, the mean value of all stellar samples is $7.3~\kmsec$ (also with some dependence on region and $\zlim$). Again, we see outliers for both $Z_\odot$ and $W_\odot$ especially for area cell B8, and to a lesser extent also for some samples in region D. In the posterior densities, there are no strong correlations between $\rho_h$ and $Z_\odot$ or between $\rho_h$ and $W_\odot$; the correlation values are typically a few percent.

In Appendix~\ref{app:control}, we show control plots for a number of stellar samples, where we compare the inferred population model with the distribution of stars in the $Z$-$w$ plane. Control plots are shown for a few representative samples, for which the population model and distribution of stars agree reasonable well. We also show the control plots for stellar samples B8-200 and B8-400, which are obviously poor fits. There is a clear over-dense feature in the $Z$-$w$ plane of the B8 stellar samples, possibly the remnant of a dissolved open cluster, which our statistical model mistakenly infers to correspond to the Galactic mid-plane. This substructure is diffuse in the space of angular coordinates, which is probably why it was not included in the open cluster catalogue of \cite{OC_catalogue} and therefore not masked. The same substructure also seems to be present, although to a lesser extent, in the neighbouring area cell B7, which exhibits similar outlier values for $Z_\odot$ and $W_\odot$.

The area cells with the largest open cluster masks are D7, D12, C10, C5, and D16 (in descending order, see Table~\ref{tab:OCmasks} for more details). Many of these stellar samples contain a lower number of stars, especially close to the mid-plane, which is essential to inferring the shape of the potential at low $|z|$. This is, at least to some extent, reflected in larger uncertainties to the inferred potential of these stellar samples (for example D7-200, D12-200, D16-200). Not only is the statistical uncertainty larger but potentially also the systematic error, as the steady state assumption is more questionable in spatial volumes where large and/or numerous open clusters are present.

In Appendix~\ref{app:Jeans}, we inferred the gravitational potential using a very simple Jeans analysis. Despite the crude nature of this analysis, it reproduced the general trends of our results: a gravitational potential that is significantly steeper than expected close to the Galactic mid-plane.

\section{Discussion}\label{sec:discussion}
We have inferred the gravitational potential of the Galactic disk for 120 stellar samples in the solar neighbourhood, in 40 separate spatial volumes, under the assumption of a steady state and separability of the gravitational potential in the vertical direction. The main reason for dividing the local volume into sub-regions was to ascertain whether spatially dependent systematic errors would affect our results. We see a spatial dependence with respect to the Galactic radius, which is consistent with a disk scale length of a few kiloparsecs. Apart from this, we see an unexpected but nonetheless clear trend, present in almost all stellar samples: The inferred vertical gravitational potential, normalised to $\Phi (0~\pc) = 0~\kmsecsq$, agrees well with the expectation at large distances from the mid-plane ($|z|\simeq 400~\pc$) but is significantly steeper than expected close to the Galactic mid-plane ($|z|\lesssim 60~\pc$). If this result is taken at face value, it implies a total matter density that is highly concentrated to the mid-plane, but decays quickly with height. This is inconsistent with the observed distribution of baryons. The high mid-plane matter density itself could potentially be explained by an excess of cold gas or other hidden matter density (see Sect.~\ref{sec:expectation} for a discussion on systematics associated with the local density of gas and Sect.~\ref{sec:intro} about the possibility of dark sector over-densities in the Galactic disk). However, even if we allow for such hidden matter densities, the low matter density inferred at greater heights ($|z| \gtrsim 300~\pc$, see Fig.~\ref{fig:A1_rho}) leaves little to no room for any matter density other than halo dark matter, which is irreconcilable with the matter density distribution and scale height of the stellar disk.

We consider time-varying dynamical effects, breaking the assumption of a steady state, to be the most probable reason for this unexpected result. We do see how local phase-space substructures can bias our result for individual stellar samples, most clearly for those of area cell B8. However, explaining the steep gravitational potential close to the Galactic mid-plane, inferred for almost all stellar samples, requires a phase-space structure that spans the whole spatial volume that is studied in this work. Indeed, spatially large time-varying phase-space structures are present in the Galaxy, for example in the form of phase-space spirals and ridges \citep{gaia_kinematics,wrinkles}, and Galactic plane mirror asymmetries (\citealt{BovyAssym}, especially prominent for heights $|z|\gtrsim 400~\pc$). In order to produce a steep gravitational potential at low $|z|$, there could be a breathing mode in the stellar disk \citep{2014MNRAS.440.1971W,2016MNRAS.457.2569M} which is currently in its most compressed state. Such a configuration would not be detectable by comparing the mean vertical velocities above and below the mid-plane, because the breathing oscillation would be at a turning point between contraction and expansion. Mass estimates close to the mid-plane and under the steady state assumption would be biased towards more massive results, as the stellar disk would have a smaller scale height and larger vertical velocities. In order to explain our results, such a breathing mode would have to be large enough for the stellar number density in the mid-plane to oscillate with a relative amplitude of $\sim 5~\%$. \citet[see for example Fig. 4]{2016MNRAS.457.2569M} have shown that a local breathing mode could be created by a spiral arm that passes close enough to the Sun, inducing a net motion away from (towards) the Galactic mid-plane for stars on the outside (inside) of the spiral arm. Furthermore, such a close passage of a spiral arm is indicated by some dynamical models of the horizontal motions within the Galactic disk (for example \citealt{2012MNRAS.425.2335S}). The dynamics of the Galactic disk are very complex and probably contains a combination of breathing and bending modes created by the last impact of the Sagittarius dwarf galaxy (for example \citealt{2019MNRAS.485.3134L}); it is very plausible to think that the famous phase-space spirals in the $z$-$w$ plane are a local manifestation of these larger scale perturbations. The phase-space spiral structure could have an effect on our analysis, especially for the stellar samples with higher $\zlim$. On a larger scale, the bending and breathing modes of the Galaxy have been shown to affect dynamical mass measurements of the Galactic disk, especially at greater heights \citep{breathingmode,Haines_2019}.

The inferred potential of our stellar samples agrees well with the results of \cite{Widmark2019}, where a similar method but different stellar sample cuts were used. Due to the smaller heights considered in that work ($|Z|<200~\pc$), there did seem to be a surplus of matter in the Galactic disk. In a similar study by \cite{Buch:2018qdr}, also using \emph{Gaia} DR2, they set an upper bound to a surface density surplus ($7.5~\Msunppcsquare$ for a scale height of $\sim 40~\pc$, 95 \% confidence region limit) that roughly correspond to the preferred model of \cite{Widmark2019}. Interestingly, the most stringent bound in \cite{Buch:2018qdr} come from their sample of A-type stars, which is a stellar population with a comparatively small scale height. The fact that their other stellar samples (G-type and F-type) produced weaker limits could possibly be explained by a time-varying dynamical structure, which could produce a different bias depending on the scale height of the stellar tracer population. The stellar samples in \cite{Widmark2019} did not include stars that were quite as bright as A-type, but still saw some anti-correlation between the inferred matter density surplus and the brightness of the stellar samples.

Apart from the overall trend of our results, we do see discrepancies larger than expected from statistical variance, especially for $Z_\odot$ and $W_\odot$. One especially noteworthy outlier is area cell B8, which appears to host a previously unknown spatially diffuse remnant of a dissolved open cluster. In this case, the population model is obviously a poor fit, as can be seen in Appendix~\ref{app:control} and Figs.~\ref{fig:control_B8-200} and \ref{fig:control_B8-400}. This demonstrates that stellar phase-space substructures are present and can produce biases, despite masking previously identified open clusters.

The shape of the inferred gravitational potential cannot be explained, at least not completely, by adding some kind of hidden matter density to the observed distribution of baryons and halo dark matter. We infer a matter density that is strongly concentrated to the Galactic mid-plane, but decays quickly with height; the baryonic distribution is inconsistent with such a rapidly decaying matter density, mainly due to the observed scale height of the stellar disk. With that said, the distribution of baryons could, to some extent, be misunderstood. As we demonstrate in Appendix~\ref{app:expectation_stars_comparison}, the baryonic model from which the expected gravitational potential is derived does not agree very well with the stellar number density profiles of our stellar samples. Most importantly, its dependence on $z$ relies on extrapolations using the different components' vertical velocity distributions, under the erroneous assumption that those velocity distributions are Gaussian (i.e. iso-thermal). It is less likely that the stellar densities in the mid-plane are significantly biased, but an update to the baryonic model could improve especially on the shape of the matter density profiles of the stellar components.

Likely, only a small bias to the results of this study could be caused by data systematics. In terms of the astrometric data of \emph{Gaia} DR2, systematic errors are probably small: $\lesssim 0.01$ mag for the apparent magnitudes \citep{photo_systematics}, $\lesssim 0.1~\mas$ for the parallax, and $\lesssim 0.1~\masyr$ for the proper motions (\citealt{astrometry_systematics}, corresponding to velocity errors $\lesssim 0.2~\kmsec$ at the distance scales of this study). Dust corrections are insignificant for the closest samples, but a poorly modeled dust map could potentially have some effect on the more distant stellar samples, especially region D. The median and 90th percentile values to the $\hat{m}_G$ dust corrections are roughly $0.02$ and $0.06$ mag for region A, and roughly $0.1$ and $0.2$ mag for region D. In summary, data and dust effects could potentially produce a small bias, but likely only for the most distant stellar samples, especially in region D. It could be worth revisiting these stellar samples with future \emph{Gaia} data releases and a more accurate dust map.

In summary, we consider the gravitational potential inferred in this work to be strongly biased by a spatially large time-varying dynamical structure, such as a breathing mode of the Galactic disk. Such a bias could potentially be diagnosed by jointly analysing a multitude of stellar populations with different dynamical properties and vertical scale heights. Another complementary measurement is explored in \cite{widmark_streams}, where they demonstrate that a cold stellar stream passing through or close to the Galactic plane could provide competitive constraints to the disk's potential, which would also be independent of the steady state assumption for the stellar disk. In a more general sense, modelling time-varying dynamical systems in order to extract information about their gravitational potential and mass distribution becomes all the more interesting and necessary, not only for weighing the Galactic disk. For the solar neighbourhood, relaxing the steady state assumption is a natural next step, and this relatively compact and well-defined problem could be a good testing ground for developing such methods.

\section{Conclusion}\label{sec:conclusion}
In this work we have inferred the gravitational potential of the Galactic disk, under the assumption of a steady state and separability in the vertical direction. We did so for 120 stellar samples, for 40 spatially separate area cells in the directions parallel to the Galactic plane, and three different spatial ranges in the vertical direction. For the inferred gravitational potentials of our stellar samples, we see a dependence on the Galactic radius that is consistent with a disk scale length of a few kiloparsecs. In addition to this, a general trend is that the gravitational potential is significantly steeper than expected close to the Galactic mid-plane ($|z|\lesssim 60~\pc$), but agrees well at greater heights ($|z|\simeq 400~\pc$). The trend is in agreement with \cite{Widmark2019}, which uses different cuts in spatial volume and absolute magnitude, and also the simple Jeans analysis carried out in Appendix~\ref{app:Jeans}. Our result does not imply a surface density surplus in the Galactic disk but rather a mass distribution that is highly concentrated to the Galactic mid-plane; this is inconsistent with the observed distribution of baryons, mainly because the observed scale height of the stellar disk cannot be reconciled with such a small total matter density at $|z|\gtrsim 300~\pc$. Our inferred gravitational potentials must be affected by a significant systematic bias; we consider a time-varying dynamical structure to be the most likely explanation. For example, our results could be explained by a breathing mode that is currently in its most compressed state, corresponding to a temporary $\sim 5~\%$ increase of the mid-plane stellar number density. This dynamical structure must also be spatially large, as it affects the full volume of all stellar samples in practically the same manner.

We will investigate the possibility of time-varying structures in future work. With future \emph{Gaia} data releases, we plan to perform the same analysis in other spatial volumes further away from the Sun, and for other stellar populations such as the brighter A-type stars studied by \cite{Schutz:2017tfp} and \cite{Buch:2018qdr}. We will also explore in more detail what kind of vibrational mode of the Galactic disk would be needed to produce our result, and how to diagnose such a feature.

Furthermore, we plan to update the baryonic matter density model in future work, using upcoming \emph{Gaia} data releases. The expected gravitational potential used in this work (as well as for example \citealt{Schutz:2017tfp}, \citealt{Sivertsson:2017rkp}, \citealt{Buch:2018qdr}, and \citealt{Widmark2019}) is based on observations of the different baryonic components' mid-plane densities and mid-plane vertical velocity distributions. As demonstrated in Appendix~\ref{app:expectation_stars_comparison}, this baryonic model could be improved using the stellar number density profiles of current or future \emph{Gaia} data releases.

\begin{acknowledgements}
We would like to thank the anonymous referee for their insightful and constructive review. AW acknowledges support from the Carlsberg Foundation via a Semper Ardens grant (CF15-0384). PFdS acknowledges support by the Vetenskapsr{\aa}det (Swedish Research Council) through contract No. 638-2013-8993 and the Oskar Klein Centre for Cosmoparticle Physics. This work made use of an HPC facility funded by a grant from VILLUM FONDEN (projectnumber 16599).

This work has made use of data from the European Space Agency (ESA) mission \emph{Gaia} (\url{https://www.cosmos.esa.int/gaia}), processed by the \emph{Gaia} Data Processing and Analysis Consortium (DPAC,
\url{https://www.cosmos.esa.int/web/gaia/dpac/consortium}). Funding for the DPAC has been provided by national institutions, in particular the institutions participating in the \emph{Gaia} Multilateral Agreement.

This research utilised the following open-source Python packages: \textsc{Matplotlib} \citep{matplotlib}, \textsc{NumPy} \citep{numpy}, \textsc{SciPy} \citep{scipy}, \textsc{Pandas} \citep{pandas}, \textsc{TensorFlow} \citep{tensorflow2015-whitepaper}, \emph{Gaia} DR2 completeness \citep{2018asclsoft11018R}.
\end{acknowledgements}




\bibliographystyle{aa} 
\bibliography{thisbib} 

\begin{appendix} 

\section{Expected baryonic density comparison}\label{app:expectation_stars_comparison}
In this section we compare the observed number density profiles of stars in regions A, B, C, and D, with the stellar density profiles of the baryonic model described in Sect.~\ref{sec:expectation}.

\begin{figure}
	\centering
	\includegraphics[width=\columnwidth]{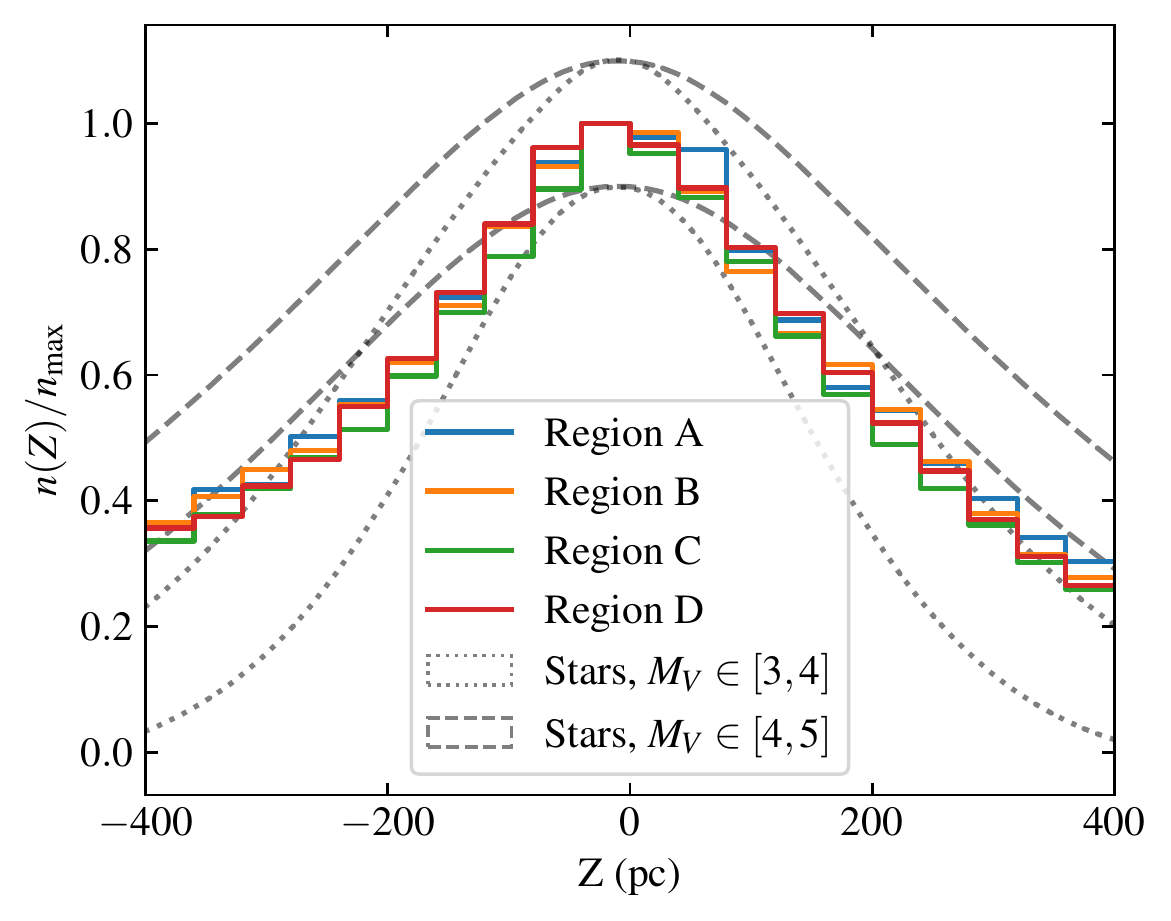}
    \caption{Stellar number densities for the stars of regions A, B, C, and D, normalised to unity for their maximum values. The number densities of two stellar components of the baryonic model are shown in terms of their $1\sigma$ bands, in dashed and dotted grey, where $Z_\odot = 10~\pc$.}
    \label{fig:comparison}
\end{figure}

This comparison is visible in Fig.~\ref{fig:comparison}. Regions A--D contain stars with different intervals in absolute magnitude $M_G$ (regions A and D have no overlap, with $I_{M_G} = [4.6,\, 5.5]$ and $I_{M_G} = [3.0,\, 4.5]$, respectively); in terms of intrinsic brightness, they correspond roughly to the baryonic model's stellar density components with absolute magnitude ranges $M_V \in [4,\, 5]$ and $M_V \in [3,\, 4]$. Evident from the figure is that neither profile of the expected density has a shape that agrees well with regions A--D. This discrepancy is explained, at least in part, by the fact that the distribution of vertical velocities is not Gaussian, as is assumed in the baryonic model, but rather a distribution with significantly heavier tails. Furthermore, the stellar regions A--D all have similar number density profiles, unlike the expected baryonic model for which the two absolute magnitude ranges have quite different profiles.

\section{HMC sampling}\label{app:sampling}
In this work, our statistical model of inference was a Bayesian hierarchical model with a high number of free parameters. Using a velocity distribution Gaussian mixture model consisting of 10 Gaussians, we had 65 degrees of freedom in the population parameters. For a stellar sample with $N$ stars, there was also a total of $7N$ degrees of freedom for the stellar parameters. However, due to the Dirac functions in the data likelihood, see Eq.~\eqref{eq:likelihood}, the degrees of freedom were effectively reduced to $4N$. Because our stellar samples each had roughly 5000 stars, the number of free parameters was typically around $2\times10^4$.

We sampled our posterior probability distribution, as described in Sect.~\ref{sec:statisticalmodel}, using a Markov chain Monte Carlo (MCMC) algorithm \citep{BayesianDataAnalysis}. In order to sample a function with such a high number of dimensions, we implemented our model in \textsc{TensorFlow} \citep{tensorflow2015-whitepaper}, which allows for auto-differentiation of the posterior density function with respect to the model's free parameters. By utilising the posterior derivatives, sampling of the posterior was made computationally tractable using Hamiltonian Monte Carlo (HMC). In addition to what is described in Sect.~\ref{sec:statisticalmodel}, we applied the constraint that the velocity dispersions $\sigma_{W,k}$ must be in ascending order, in order to avoid multiplicity in our velocity Gaussian mixture model.

Before running our MCMC chains, we ran an initial minimisation algorithm for each stellar sample, in order to locate the approximate mode of the population model. During this minimisation the stellar parameters were fixed and only the velocity information of stars with high data quality were used (meaning $\hat{\sigma}_{\varpi}<0.2~\mas$ and $\hat{\sigma}_{\text{RV}}<2~\kmsec$). In order to avoid getting stuck in a local minimum far from the global minimum, this minimisation was run from 10 randomly chosen initial points in the space of the population parameters, out of which the minimum posterior value was chosen.

After this, we ran a thorough burn-in phase for the full posterior probability density, where its mode was located and the diagonal step-size matrix of the HMC algorithm was tuned. We then ran the chain for long enough to obtain a minimum of 2500 independently drawn posterior realisations for each stellar sample. This was tested by calculating the auto-correlation of each individual free parameter of our model. For each free parameter, we counted the number of alternating crossings of the 16th and 84th percentiles of the obtained posterior density, in order to ensure that the full parameter space was well sampled.

The \textsc{TensorFlow} code written for this work is open source and can be found online (\url{https://github.com/AxelWidmark/mosaik}).

\section{Comparison with Jeans analysis}\label{app:Jeans}

The statistical method used in this work is rather complicated, and yields an unexpected result. In this section we present a similar but significantly simpler analysis, which used an easily reproducible method based on the Jeans equations. Despite the crude nature of this approach, it reproduces the general trend of our results.

Using the assumption of separability of the gravitational potential, reducing the problem to the vertical dimension only, the vertical Jeans equation states that
\begin{equation}\label{eq:Jeans}
    \frac{1}{n}\frac{\partial}{\partial Z}\left(n \, \sigma_W^2 \right) + \frac{\partial \Phi}{\partial Z} = 0,
\end{equation}
where $\sigma_W^2$ refers the variance of the vertical velocity distribution.

We applied the Jeans analysis to the four regions A, B, C, and D (defined by the rings formed by area cells A1--A4, B1--B8, C1--C12, and D1--D16), with a height limit of $\zlim = 400~\pc$ above and below the Sun. We did not apply any masks for the open clusters and accounted for no \emph{Gaia} DR2 incompleteness effects or observational uncertainties.

Each respective region was divided into bins with a height of 80 pc. For each bin, labelled with an index $l$, the stellar number density $n_l$ was given by the number of stars in that bin, and the variance $\sigma_{W,l}^2$ was calculated from the stars in that bin with sufficient data quality (requiring that $\hat{\sigma}_{\text{RV}} < 2~\kmsec$ and $\hat{\sigma}_\varpi < 0.2~\mas$). By using a discretised version of Eq.~\eqref{eq:Jeans}, the difference in the gravitational potential of two neighbouring bins with indices $l+1$ and $l$ is equal to
\begin{equation}\label{eq:Jeans_discrete}
    \Phi_{l+1}-\Phi_{l} =
    -\frac{2}{n_{l+1}+n_{l}} \left(n_{l+1} \sigma_{W,l+1}^2 - 
    n_{l} \sigma_{W,l}^2\right).
\end{equation}
For the inferred gravitational potential of each respective bin, we estimated the statistical error by jackknifing. The number of stars in each respective region is presented in Table~\ref{tab:jeans_regions}.
{\renewcommand{\arraystretch}{1.6}
\begin{table}
\caption{Number of stars in the regions used for Jeans analysis. The columns show: region; the total number of stars, used to calculate the stellar number density; and the number of stars with good quality data ($\hat{\sigma}_{\text{RV}} < 2~\kmsec$ and $\hat{\sigma}_\varpi < 0.2~\mas$), used to calculate the velocity dispersion.}
\label{tab:jeans_regions}      
\centering          
\begin{tabular}{l l l} 
\hline
Region & Total number & Good quality \\
\hline
A & 29,395 & 21,093 \\
B & 59,240 & 44,833 \\
C & 89,122 & 67,152 \\
D & 96,005 & 71,293 \\
\hline                  
\end{tabular}
\end{table}}

\begin{figure}
	\centering
	\includegraphics[width=\columnwidth]{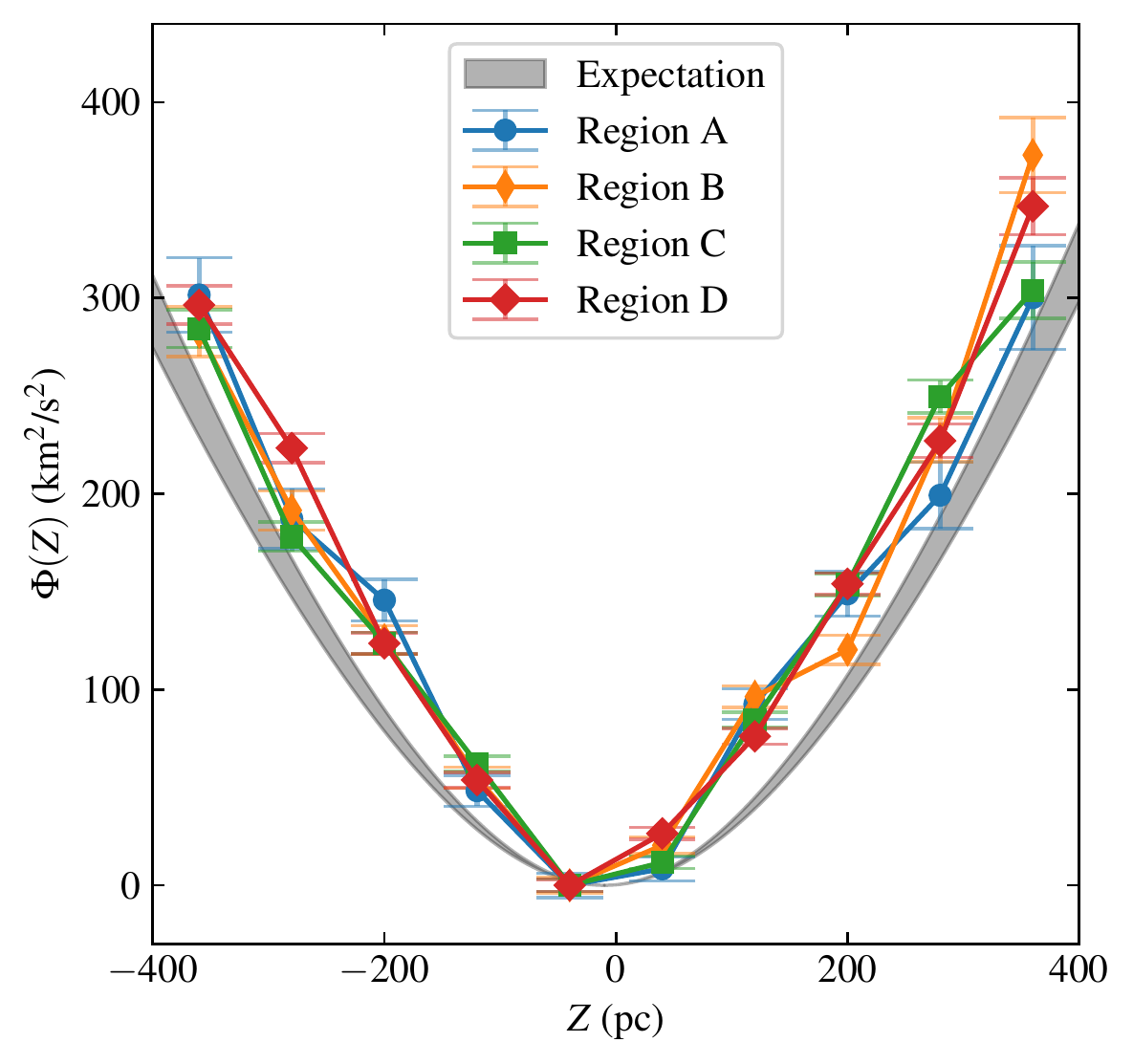}
    \caption{Inferred gravitational potential using the simple Jeans analysis, for regions A--D. The grey line represents the expected case described in Sect.~\ref{sec:expectation}, where $Z_\odot = 10~\pc$. The errorbars represent $1\sigma$ statistical uncertainties.}
    \label{fig:Jeans}
\end{figure}

The inferred gravitational potentials of regions A--D are visible in Fig.~\ref{fig:Jeans}, where they are normalised such that that their minimum values are equal to zero. The inferred potential is close to the expectation at higher values of $|z|$, but is significantly steeper close to the mid-plane, in agreement with the main analysis of this work. The inferred potential becomes increasingly noisy if choosing smaller bins ($<80~\pc$), especially for region A which has comparatively few stars, but still tends to the same overall result.

\section{Control plots}\label{app:control}
In this section, we compare the inferred population model with the distribution of stars, in terms of their phase-space distributions in vertical position and vertical velocity. The difference in the stellar number counts between the two are quantified in terms of a residual in the $Z$-$w$ plane. It is defined
\begin{equation}
    \text{residual} = \frac{\text{stars}-\text{model}}{\sqrt{\text{model}}},
\end{equation}
where `model' refers to the number of stars in a pixel as predicted by the population model, and `stars' refers to the number of stars according to the stellar parameters. The latter could in principle be called `data', as the phase-space position is strongly constrained by the data for the majority of stars in our stellar samples. In the limit of high numbers, the residuals should be distributed according to a Gaussian with a standard deviation of one; however, because some pixels have a small number count, especially at the sample boundary around $|Z|\simeq\zlim$, the distribution of residuals is skewed.

\begin{figure*}
	\centering
	\includegraphics[width=0.85\textwidth]{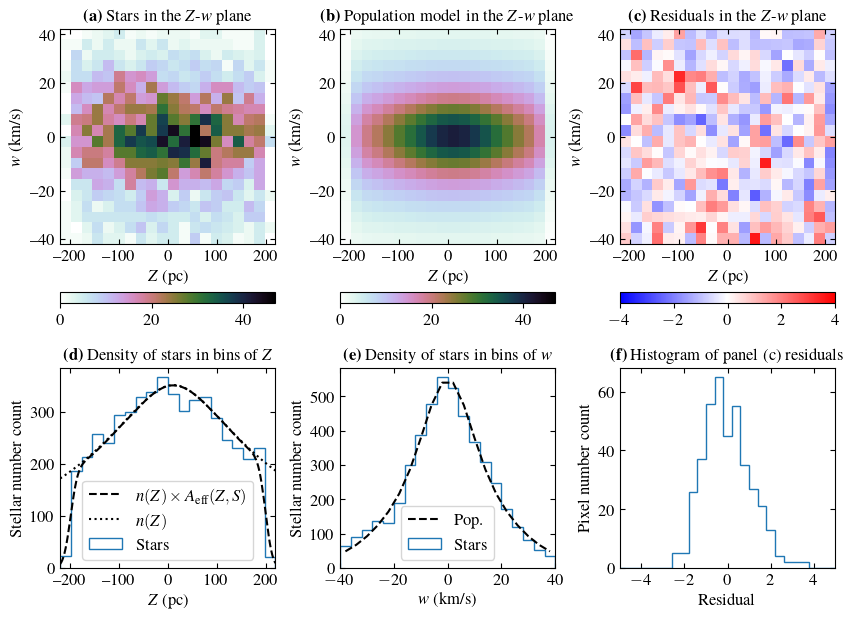}
    \caption{Control plot for sample A1-200, showing the distribution of vertical position $Z$ and vertical velocity $w$ in one-dimensional and two-dimensional histograms, for a randomly drawn realisation of the posterior probability density. The colour bars of panels (a) and (b) show the number of stars in each respective pixel in the $Z$-$w$ plane, and use the same normalisation. The colour bar in panel (c) shows the residuals of each respective pixel, which is the tension between the phase-space number count according to the population model and the stellar parameters (which is strongly constrained by the data). In panel (d), the dashed and dotted black lines correspond to the stellar number density according to the population model, where the former includes the effective area. Similarly, the dashed black line in panel (e) shows the distribution of vertical velocities (marginalised over $Z$) according to the population model.}
    \label{fig:control_A1-200}
\end{figure*}

\begin{figure*}
	\centering
	\includegraphics[width=0.85\textwidth]{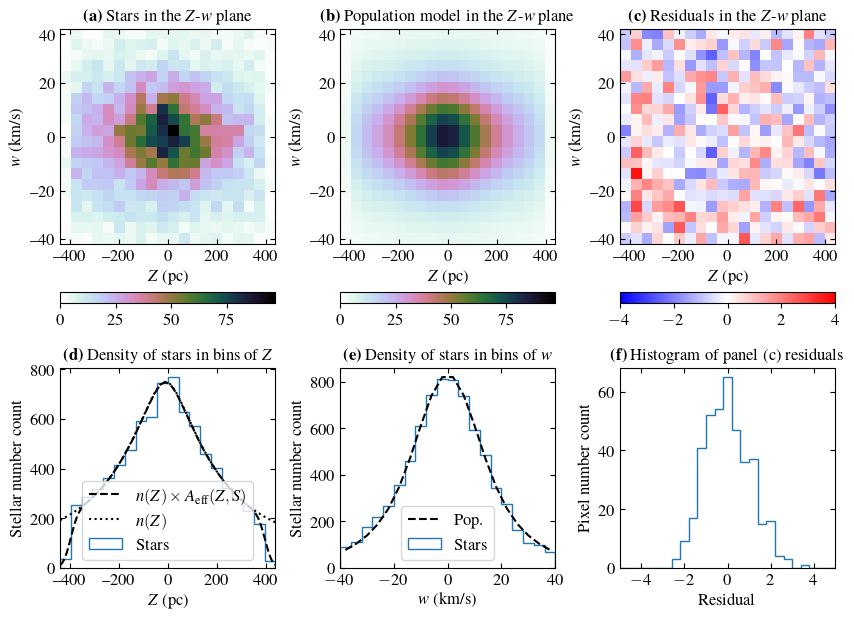}
    \caption{Same as Fig.~\ref{fig:control_A1-200}, but for stellar sample C1-400.}
    \label{fig:control_C1-400}
\end{figure*}

\begin{figure*}
	\centering
	\includegraphics[width=0.85\textwidth]{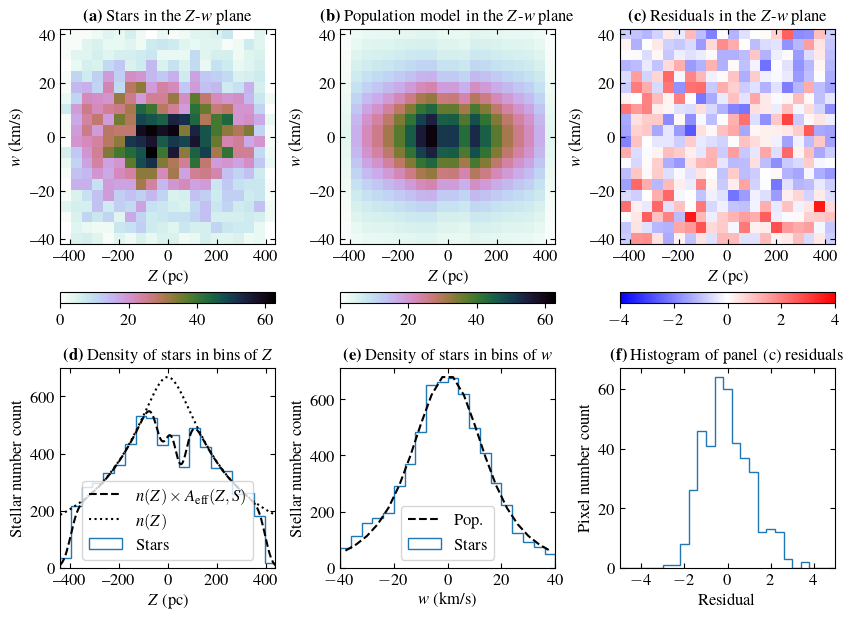}
    \caption{Same as Fig.~\ref{fig:control_A1-200}, but for stellar sample C5-400.}
    \label{fig:control_C5-400}
\end{figure*}

\begin{figure*}
	\centering
	\includegraphics[width=0.85\textwidth]{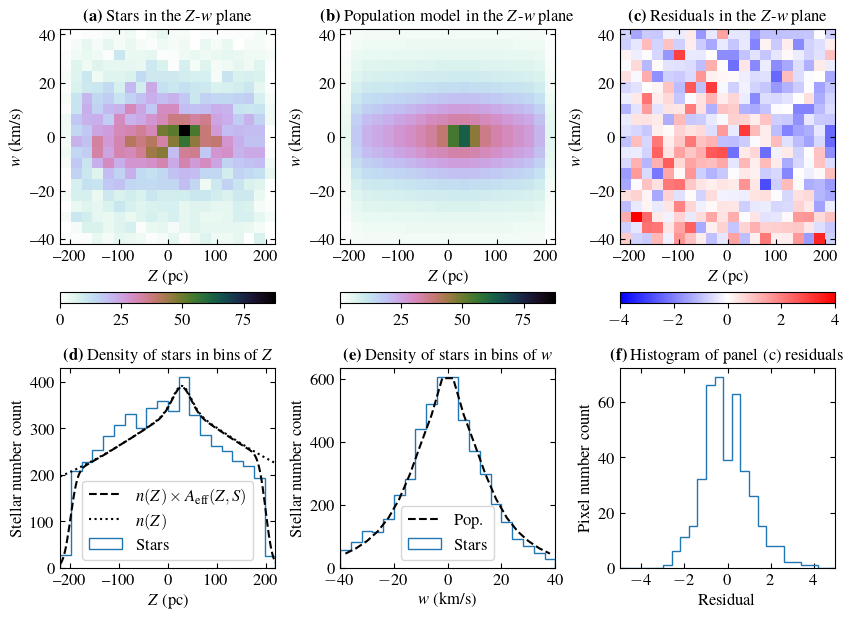}
    \caption{Same as Fig.~\ref{fig:control_A1-200}, but for stellar sample B8-200.}
    \label{fig:control_B8-200}
\end{figure*}

\begin{figure*}
	\centering
	\includegraphics[width=0.85\textwidth]{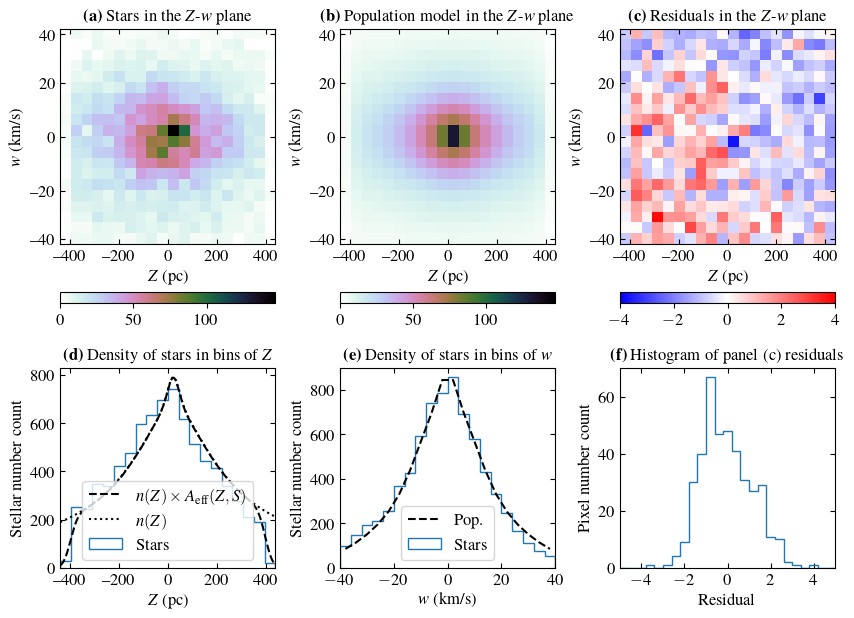}
    \caption{Same as Fig.~\ref{fig:control_A1-200}, but for stellar sample B8-400.}
    \label{fig:control_B8-400}
\end{figure*}

Control plots are shown in Figs.~\ref{fig:control_A1-200}--\ref{fig:control_B8-400}, for samples A1-200, C1-400, C5-400, B8-200, and B8-400. The first three are included as representative examples, where the distribution of stars in $Z$ and $w$ are reasonably well fitted by the population model. There is some structure that our model does not capture, for example in the form of slight asymmetries with respect the Galactic plane. Apart from this asymmetry there are no obvious phase-space features that are shared between all area cells. Stellar sample C5-400 is an example of a sample with significant open cluster masks. The latter two, B8-200 and B8-400, are special cases. As discussed in Sect.~\ref{sec:results}, area cell B8 has a substructure in the $Z$-$w$ plane which our model mistakenly infers to correspond to the Galactic mid-plane, and Figs.~\ref{fig:control_B8-200} and \ref{fig:control_B8-400} clearly shows that the inferred phase-space distributions are poor fits.

The control plots of all 120 stellar samples can be found online (\url{https://github.com/AxelWidmark/mosaik/tree/main/AllControlPlots}).

\end{appendix}

\end{document}